\documentclass[aps,floats,twocolumn,epsf,prl,showpacs]{revtex4-1}
\usepackage{graphicx}
\usepackage{epstopdf}
\usepackage{amsmath}
\usepackage{amssymb}
\usepackage[colorlinks=true,urlcolor=blue,citecolor=blue,linkcolor=blue,breaklinks=true]{hyperref}
\usepackage{color}
\usepackage[normalem]{ulem}
\usepackage[dvipsnames]{xcolor}

\usepackage{ulem}

\newcommand{\pw}[1]{{\color{black}#1}}
\newcommand{\kh}[1]{{\color{black}#1}}
\newcommand{\mk}[1]{{\color{black}{#1}}}

\begin{document}

\title{Optimizing superconductivity: from cuprates via nickelates to palladates}

\author{Motoharu Kitatani$^{a,b}$, Liang Si$^{c,d}$, Paul Worm$^d$, Jan M. Tomczak$^{d,e}$, Ryotaro Arita$^{b,f}$ and Karsten Held$^d$}

\affiliation{$^a$Department of Material Science, University of Hyogo, Ako, Hyogo 678-1297, Japan}
\affiliation{$^b$RIKEN Center for Emergent Matter Sciences (CEMS), Wako, Saitama, 351-0198, Japan}
\affiliation{$^c$School of Physics, Northwest University, Xi’an 710127, China}
\affiliation{$^d$Institute of Solid State Physics, TU Wien, 1040 Vienna, Austria}
\affiliation{$^e$Department of Physics, King’s College London, Strand, London WC2R 2LS, United Kingdom}
\affiliation{$^f$Research Center for Advanced Science and Technology, University of Tokyo 4-6-1, Komaba, Meguro-ku, Tokyo 153-8904, Japan
}

\date{\today}

\begin{abstract}
  Motivated by cuprate and nickelate superconductors, we perform a comprehensive study of the superconducting instability in the single-band Hubbard model. We calculate the  spectrum and superconducting transition temperature $T_{\rm c}$ as a function of filling and Coulomb interaction for a range of hopping parameters, using the dynamical vertex approximation. We find the sweet spot for high $T_{\rm c}$ to be at intermediate coupling, moderate Fermi surface warping, and low hole doping. Combining these results with first principles calculations, neither nickelates nor cuprates are close to this optimum within the single-band description. Instead, we identify some palladates, notably RbSr$_2$PdO$_3$ and $A^{\prime}_2$PdO$_2$Cl$_2$ ($A^{\prime}$=Ba$_{0.5}$La$_{0.5}$), to be virtually optimal, while others, such as NdPdO$_2$, are too weakly correlated.
\end{abstract}

\maketitle

\noindent 
{\sl Introduction}---Ever since the discovery of cuprate superconductivity~\cite{Bednorz1986}, the material dependence of the transition temperature $T_{\rm c}$ and exploring routes toward optimizing $T_{\rm c}$ are a central quest of condensed matter physics. 
The recently discovered nickelate superconductors provide a new perspective to this quest. Similar to cuprates, superconductivity in nickelates emerges from a doped 3$d^{9-\delta}$ ($\delta\sim$0.2) electronic configuration of the transition metal. Besides the initial infinite-layer superconductor Nd$_{1-x}$Sr$_x$NiO$_2$ \cite{li2019superconductivity,Si2019,Werner2019,zeng2020,Li2020,Hepting2020,Nomura2022}, substituting  neodymium with another lanthanoid \cite{Osada2020,Zeng2021,Osada2021} and also the quintuple-layer compound \cite{pan2022} show superconductivity. This indicates that, akin to cuprates, there is a whole family of nickelate superconductors.

\kh{
As for the theoretical modeling, the one-band Hubbard model is arguably the simplest effective  model for cuprates \cite{Anderson1987,zhang1988effective}.
Its tight-binding parameters can be
obtained from {\em ab initio} calculations, and 
the relation between model parameters and the experimental $T_{\rm c}$ has been analyzed
\cite{Pavarini2001,Sakakibara2010,Weber2012,Sakakibara2014,Jang2016,Teranishi2018,Hirayama2018,Nilsson2019,Sakakibara2019,Hirayama2019,Teranishi2021,Watanabe2021,Moree2022}.
For nickelates, a similar scenario (1) with a one-band Hubbard model plus largely decoupled electron pockets has been
put forward  \cite{Kitatani2020,Karp2020,Worm2022,Xie2022,Chen2022,10.3389/fphy.2021.810394,Karp2022,Lane2022arXiv}. 
Based on the same density functional theory (DFT) and dynamical mean-field theory (DMFT) Fermi surface with
Ni $3d_{x^2-y^2}$ orbital  plus electron pockets
around $A$ and $\Gamma$ momentum, a second group of scenarios (2) \cite{Lee2004,Adhikary2020,Wang2020,WangZ2020}
emphasizes the role of holes in the  Ni $3d_{z^2}$ orbital. 
\mk{These originate from an admixture around the  $\Gamma$ pocket that is predominately  Nd $5d_{z^2}$.
In} \pw{ scenario (1) this is argued not to be of primary importance for superconductivity because of the strong doping and rare earth cation dependence of the 
$\Gamma$ pocket \cite{Si2019,PhysRevB.101.241108,Ryee2019,Been2021}.}
Finally, scenario (3) proposes an {\em additional}
 Ni $3d_{z^2}$ Fermi surface based on
self-interaction corrected (sic) DFT+DMFT
\cite{Lechermann2019,Lechermann2020,Kreisel2022,Lechermann2022,Chen2022}.
Such an additional Fermi surface is also obtained in  antiferromagnetically ordered DFT  \cite{Wan2021,MiYoung2020}, $GW$+DMFT \cite{Petocchi2020}, and DFT+DMFT in the overdoped region \cite{Kitatani2020,pockets}.
}

\kh{
While the relevant low energy model for nickelates is still under debate, a boost for scenario (1) was its successful  prediction of the  superconducting phase diagram \cite{Kitatani2020} prior to experiments~\cite{zeng2020,Li2020} and with
\pw{high} accuracy in the light of 
new,  defect-free films \cite{Lee2022}. Also some other experiments including, among others, the Hall coefficient, resonant x-ray spectroscopy  \cite{Higashi2021} and magneto transport \cite{sun2022evidence} point toward this scenario. As for the microscopic origin of high-$T_c$ superconductivity: while spin fluctuations mediate superconductivity in Ref.~\cite{Kitatani2020}, the topic remains highly controversial; many different mechanisms have been proposed \cite{Lee2006,Scalapino2012,Fradkin2015,Keimer2015}.}

\kh{
The aim of the present paper is hence two-fold: First, we would like to identify the optimal conditions for superconductivity in the Hubbard model building upon recent progress made with diagrammatic extensions of DMFT \cite{Rohringer2018}. In particular, we will employ the dynamical vertex approximation (D$\Gamma$A)  \cite{Toschi2007}, which accurately describes antiferromagnetic spin fluctuations in the parameter range where numerical quantum Monte-Carlo simulations are still available \cite{Schaefer2021}. Second, from a materials point of view we would like to identify 
cuprate- or nickelate-like materials that promise even higher $T_c$'s.  
One important factor is the interaction strength  $U$ and its ratio to the hopping $U/t$. On a qualitative level it has  been recently found   \cite{Sakakibara2020,Kitatani2020} that  the interaction strength is too large in nickelates. Higher $T_c$'s should thus be possible using compressive strain  or pressure  \cite{cryst12050656}, confirmed experimentally by a record $T_{\rm c}>30\,$K for nickelates under a pressure of 12 GPa with no saturation yet discernible \cite{Wang2021}.  Much more dramatic changes of $U/t$ are possible when going from  3$d$ to 4$d$ transition metal oxides, which can be achieved by replacing Ni with Pd \cite{PhysRevMaterials.2.104803,10.3389/fphy.2021.810394,Motoaki2019,Kitatani2020}.
Here, we study this possibility on a quantitative level. We show that a one-band description is justified for palladates, and make a prediction of the superconducting phase diagram
which can be tested in experiment.
}

\begin{figure}[tb]
        \centering
         \includegraphics[width=.8\linewidth,angle=0]{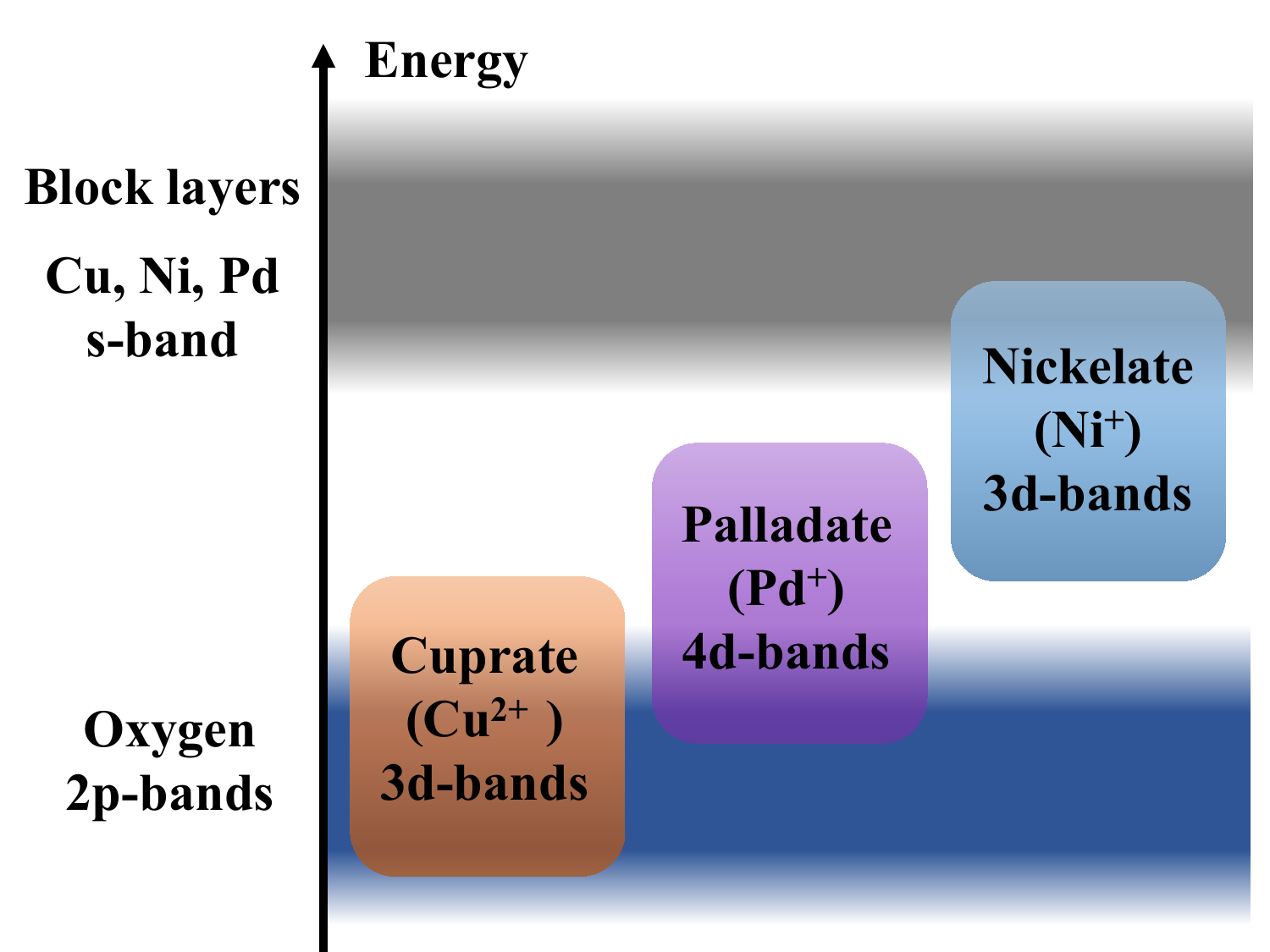}
\caption{(Color online) Schematic picture of the energy levels for copper (Cu$^{2+}$), nickel (Ni$^+$), palladium (Pd$^+$) superconductors. 
        \label{Fig:elements}}
\end{figure}
\kh{Let us start by sketching} the electronic structures of the above-mentioned materials schematically in Fig.~\ref{Fig:elements}. As is well known, the parent cuprate compounds are charge transfer insulators \cite{Zaanen1985};  Cu-3$d$ and O-$2p$ bands exist at a similar energy level and are strongly hybridized. Thus the Hubbard model is justified only as an effective model that mimics the physics of the Zhang-Rice singlet \cite{zhang1988effective}. In contrast, the Ni 3$d$-orbital is higher up in energy by $\sim$1\,eV. 
The increased distance to the oxygen orbitals then makes the single-$d_{x^2-y^2}$-orbital description more suitable than for the CuO$_2$ layers in cuprates. However, density-functional theory (DFT) calculations  \cite{Botana2019,Sakakibara2020,Motoaki2019,hu2019twoband,Wu2019,Nomura2019,Zhang2019,Jiang2019} show that the Ni-$d$ bands of NdNiO$_2$ now  partially overlap \kh{instead} with the bands of Nd  in-between the NiO$_2$ layers, forming \kh{the already mentioned} pockets around the $A$ and \mk{$\Gamma$ momentum.} 

If we replace Ni ($3d$) with Pd ($4d$), the Pd $4d$-orbitals are shifted back down by $\sim$1\,eV \kh{ due to the higher ionization energy of Pd compared to Ni. This removes the pockets present in the nickelates and leads to a larger  $d$-$p$ hybridization and minor overlap with the oxygen bands. However, we do not yet have a  charge transfer state as in the cuprates.}
Indeed, our DFT \kh{and DMFT} computations (shown in Supplementary Materials (SM) \cite{SM} Section~I-\kh{III}) of the crystal and electronic structures of the nickelate NdNiO$_2$ and the palladates NdPdO$_2$, RbSr$_2$PdO$_3$ and $A^\prime_2$PdO$_2$Cl$_2$ show that palladate compounds are somewhere in-between cuprates and nickelates -- with a single-band $d_{x^2-y^2}$ Fermi surface \kh{justifying a modelling by a single-orbital Hubbard model}. Tuning the dispersion and interaction strength in palladates \kh{thus} opens so-far untapped possibilities for finding new superconductors with possibly higher $T_{\rm c}$'s.

\noindent
{\sl Model and Method}---We study the two-dimensional Hubbard model on the square lattice with Hamiltonian 
\begin{align}
    {\cal H}= \sum_{\mathbf{k},\sigma}
    \epsilon_{\mathbf{k}} c^{\dag}_{\mathbf{k},\sigma} c_{\mathbf{k},\sigma}^{\phantom{\dag}}
    +U \sum_{i} n_{i,\uparrow}n_{i,\downarrow},
\end{align}
where $c^{\dag}_{\mathbf{k},\sigma}(c_{\mathbf{k},\sigma}^{\phantom{\dag}})$ is the creation (annihilation) operator,
\begin{align}
    \epsilon_{\mathbf{k}}=
    -2t[{\rm cos}(k_x)+{\rm cos}(k_y)]
    -4t^{\prime}{\rm cos}(k_x){\rm cos}(k_y) \nonumber \\
    -2t^{\prime\prime}[{\rm cos}(2k_x)+{\rm cos}(2k_y)],
\end{align}
the energy-momentum dispersion, $U$ the onsite Coulomb repulsion, and $t,t^{\prime},t^{\prime\prime}$ are the nearest, second nearest, and third nearest hoppings, respectively. The model parameters are obtained from \kh{DFT},
using  \textsc{wien2k}  \cite{wien2k2020} 
with the PBE \cite{PhysRevLett.77.3865,PhysRevLett.100.136406}
exchange correlation functional and  \textsc{wien2wannier} \cite{kunevs2010wien2wannier} for projecting onto a maximally localized 3$d_{x^2-y^2}$ Wannier orbital \cite{RevModPhys.84.1419}.
\kh{SM Sec. VI, VII, VIII, and IX provides  details on the Wannier projection, the DFT calculation of $A^{\prime}_2$PdO$_2$Cl$_2$,  the  stability against structural distortions, and the antiferromagnetic DFT solution, respectively.}
\kh{Before} constructing the single-orbital model for palladates, we performed multi-orbital DMFT calculations \kh{which confirm} the single orbital nature of the system, see SM \cite{SM} Sec. III. The constrained random phase approximation (cRPA) is employed to estimate $U$. 
Following the previous research \cite{Kitatani2020}, we employed slightly enhanced values (+0.35\,eV \cite{Uenhancement}) from our cRPA calculation for entangled bands \cite{Miyake2009}: 2.85\,eV for NdNiO$_2$, 2.55\,eV for RbSr$_2$PdO$_2$, and 2.97\,eV for $A^{\prime}_2$PdO$_2$Cl$_2$ (which are consistent with the preceding study \cite{Motoaki2019}; please note that small changes in the interaction strength $U$ do not change our conclusion). Table~\ref{tab:my_label} provides a summary of the DFT and cRPA derived parameters \cite{Pd-parameters} (for details see SM \cite{SM} Section~II), which are used in subsequent D$\Gamma$A calculations. Besides these material-specific Hubbard models we also include  the simplest case with nearest-neighbor hopping only ($t^{\prime}=t^{\prime\prime}=0$).

\begin{table}[tb]
    \centering
    \begin{tabular}{c|cccc}
        \hspace{30mm} & \hspace{2mm}$|t|({\rm meV})$\hspace{2mm} & \hspace{1mm}$t^{\prime}/t$\hspace{1mm} & \hspace{1mm}$t^{\prime\prime}/t$\hspace{1mm} & \hspace{1mm}$U_{\rm eff}/t$\hspace{1mm} \\
        \hline \hline
        NdNiO$_2$ & 395 &  -0.25 & 0.12 & 8 \\
        NdNiO$_2$(Strained) & 419 & -0.23 & 0.12 & 7.0--7.5  \\
        NdPdO$_2$ & 558 &  -0.17 & 0.13 & 4.5 \\
        RbSr$_2$PdO$_3$ & 495 &  -0.24 & 0.16 & 6 \\
        $A^{\prime}_2$PdO$_2$Cl$_2$  & 443 &  -0.22 & 0.14 & 7.5 \\
        $A^{\prime}_2$PdO$_2$Cl$_2$ (-1.5\% strain)  & 470 &  -0.22 & 0.14 & 7.0 \\
        $A^{\prime}_2$PdO$_2$Cl$_2$ (-3.0\% strain) & 497 &  -0.22 & 0.14 & 6.0 \\
        \hline
    \end{tabular}
    \caption{Summary of the DFT-derived parameters for the single-band Hubbard model, as an effective low-energy model for the nickelate NdNiO$_2$, the palladates NdPdO$_2$, RbSr$_2$PdO$_3$, and $A^{\prime}_2$PdO$_2$Cl$_2$. 
    }
    \label{tab:my_label}
\end{table}

We analyze these  single-orbital models by means of the \kh{D$\Gamma$A} \cite{Toschi2007,held2008,Katanin2009,Kusunose2006},  a diagrammatic extension of \kh{DMFT} \cite{PhysRevLett.62.324,RevModPhys.68.13,kotliar2004strongly}. Similar techniques have been previously applied to 
unconventional superconductivity on the square lattice \cite{Otsuki2014,Kitatani2015,Kitatani2017,Vucicevic2017,Vilardi2019,Sayyad2020,Astretsov2020,Kitatani2022arXivAPL}.
We mainly use the continuous-time quantum Monte-Carlo solver from w2dynamics\cite{wallerberger2019w2dynamics} as DMFT solver, see SM \cite{SM} for details.
D$\Gamma$A simultaneously includes strong correlations  and  long-range spatial (charge and spin) fluctuations.\
Both are essential for modelling superconductivity in correlated electron systems. Most importantly, D$\Gamma$A can describe dynamical screening effects which are crucial for accurately determining $T_{\rm c}$ \cite{Kitatani2019}. It has predicted the superconducting dome in nickelates \cite{Kitatani2020} prior to experiments~\cite{zeng2020,Li2020,Lee2022} with astonishing accuracy. For a review of D$\Gamma$A, see \cite{Rohringer2018}; and \cite{Kitatani2022} for how to calculate $T_{\rm c}$.

\begin{figure}[t]
        \centering
                \includegraphics[width=\linewidth,angle=0]{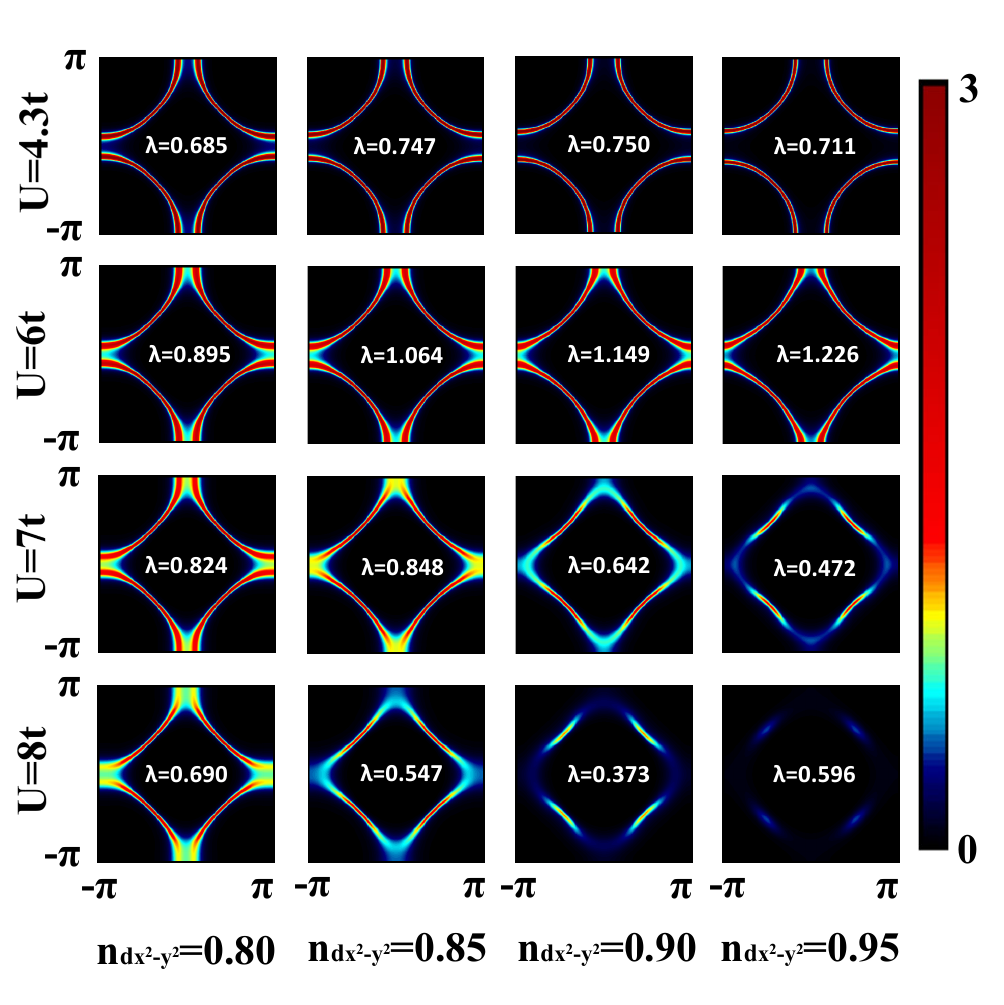}
        \caption{(Color online) D$\Gamma$A Spectrum (momentum dependence of the imaginary part of the Green function $-\Im G/\pi$ at lowest Matsubara frequency) for different interactions ($U/t=4.3,6.0,7.0,8.0$) and fillings ($n=0.80,0.85,0.90,0.95$) at $T=0.01t,t^{\prime}/t=-0.17,t^{\prime\prime}/t=0.13$. Corresponding $d$-wave superconductivity eigenvalues $\lambda$ are also shown. 
        \label{Fig:spctrum}}
\end{figure}

\noindent
{\sl Spectrum}---We first discuss the electronic spectrum. In Fig.~2, we show the momentum dependence of the imaginary part of the Green's function at the lowest Matsubara frequency:  $-\Im G(\omega_n=\pi/\beta,\mathbf{k})/\pi$ for \mk{various interactions:} $U/t=4.3,6,7,8$ \mk{and fillings:}
$n=0.80,0.85,0.90,0.95$ for hoppings corresponding to NdPdO$_2$ ($t^{\prime}/t = -0.17,t^{\prime\prime}/t = 0.13$). We observe that the spectrum changes from a non-interacting-like Fermi surface at weak coupling to a shape with strongly momentum-dependent damping at stronger coupling, before, finally, all spectral weight is removed. In between, we obtained a Fermi arc structure at low doping ($n_{d_{x^2-y^2}}\sim0.90-0.95$) for $U=7t$ and $8t$, which is a hallmark feature of cuprates.

Even outside the pseudogap region, correlation effects change the Fermi surface structure. Specifically, they decrease the Fermi surface warping, i.e. effectively decrease $t'$. 
Such change of the spectrum for large $(t^{\prime},t^{\prime\prime})$ region has also been observed in other theoretical studies \cite{Wu2018,Rossi2020,Klett2022}. 
These Fermi surface changes also affect the superconducting instability. While this effect (coming from the momentum dependence of $\Re \Sigma$) is minor compared with the pseudogap physics (stemming from the momentum dependence of $\Im \Sigma$), we find that such a Fermi surface flattening can enhance superconductivity, but only around the optimal conditions in parameter space; see SM \cite{SM} for a detailed discussion.

As demonstrated here, D$\Gamma$A properly captures correlation induced changes of the Fermi surface (e.g., Fermi arc in cuprates \cite{Yoshida2006}) and is consistent with previous results.

\begin{figure}[tbp]
        \centering
                \includegraphics[width=\linewidth,angle=0]{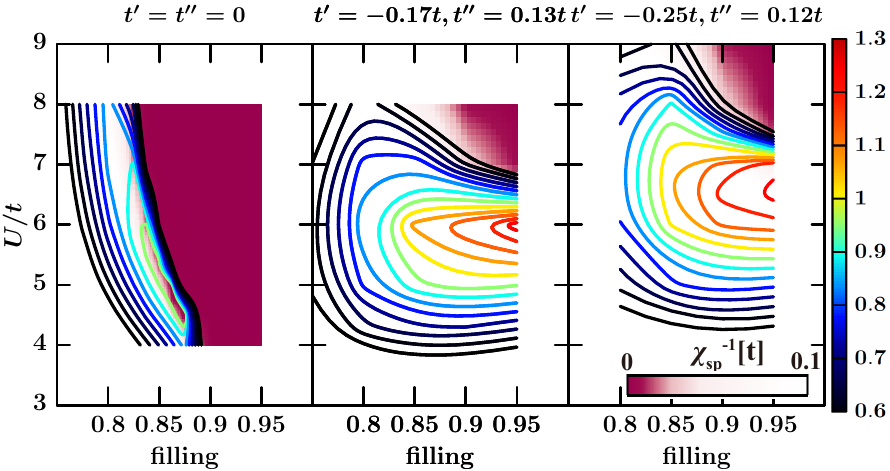}
        \caption{(Color online) Superconducting eigenvalue $\lambda$ and antiferromagnetic susceptibility $\chi_{\rm sp}{(Q_{\rm max},\omega=0)}$ as a function of interaction $U$ and filling at $T=0.01t$  for the three different  $t'$, $t''$ given at the top of each panel.
        \label{Fig:lambda}}
\end{figure}

\noindent
{\sl Superconductivity}---Next, we discuss the superconducting instability. To do so, we calculate the eigenvalues of the linearized gap (Eliashberg) equation which is the usual procedure for evaluating the superconducting instability from the paramagnetic solution. The eigenvalue $\lambda$ is a measure of the superconducting instability, and $T_{\rm c}$ is identified by $\lambda$ reaching unity. While D$\Gamma$A is unbiased with respect to spin, charge and quantum critical fluctuations, we find that spin fluctuations mediate $d$-wave superconductivity in all cases studied.

In Fig.~\ref{Fig:lambda}, we plot the superconducting eigenvalues against the interaction $U$ and the filling for three tight-binding parameter sets: the simplest case ($t^{\prime}=t^{\prime\prime}=0$), parameters for NdPdO$_2$ ($t^{\prime}=-0.17t,t^{\prime\prime}=0.13t$) and NdNiO$_2$ ($t^{\prime}=-0.25t,t^{\prime\prime}=0.12t$). First, we notice in all cases a strong suppression of $\lambda$ around half-filling at strong coupling.
This leads to a dome structure of $T_{\rm c}$ as a function of both interaction strength and filling, which is essential for optimizing superconducting materials.

Fig.~\ref{Fig:lambda}  unequivocally  reveals that the origin of this suppression is too strong 
antiferromagnetic correlations (dark red color scale).
These open a pseudogap  in Fig.~\ref{Fig:spctrum} and thus  suppress the electron propagator. Even though the antiferromagnetic pairing glue is huge, this eventually suppresses superconductivity.

Let us note that cluster DMFT \cite{Chen2015} shows a similar tendency as in Fig.~\ref{Fig:lambda}, albeit with still
much higher temperatures and weaker superconducting fluctuations. The pseudogap behavior is, on the other hand, consistent with the observation that the insulating regime expands if long-range spatial fluctuations are properly included \cite{Schaefer2015,Simkovic2020,Schaefer2021}.

Except for the perfectly nested case ($t^{\prime}=t^{\prime\prime}=0$) where spin-fluctuations are overly strong and open a gap regardless of interaction strength, we can separate the $U$-range into two regions: For weak couplings ($U \lesssim 5$), there is only a weak doping dependence. For strong coupling ($U\gtrsim 7$), on the other hand, a pronounced filling dependence develops, with antiferromagnetic fluctuation dominating around half-filling and suppressing $\lambda$. Optimal conditions for superconductivity in the one-band 2D square-lattice Hubbard model are realized in between these two regions.
The importance of the finite Fermi surface warping was also suggested in early phenomenological material-dependence studies of cuprate superconductors \cite{Pavarini2001}.
The only other possibilities to  significantly enhance $T_{\rm c}$ are (i) increasing $t$, which sets the energy scale, while keeping all parameter ratios constant, or (ii) creating a positive feedback on superconductivity from other bands including the oxygen bands in case of cuprates \cite{Weber2012,Rybicki2016,Kowalski2021}.

\begin{figure}[tbp]
        \centering
                \includegraphics[width=\linewidth,angle=0]{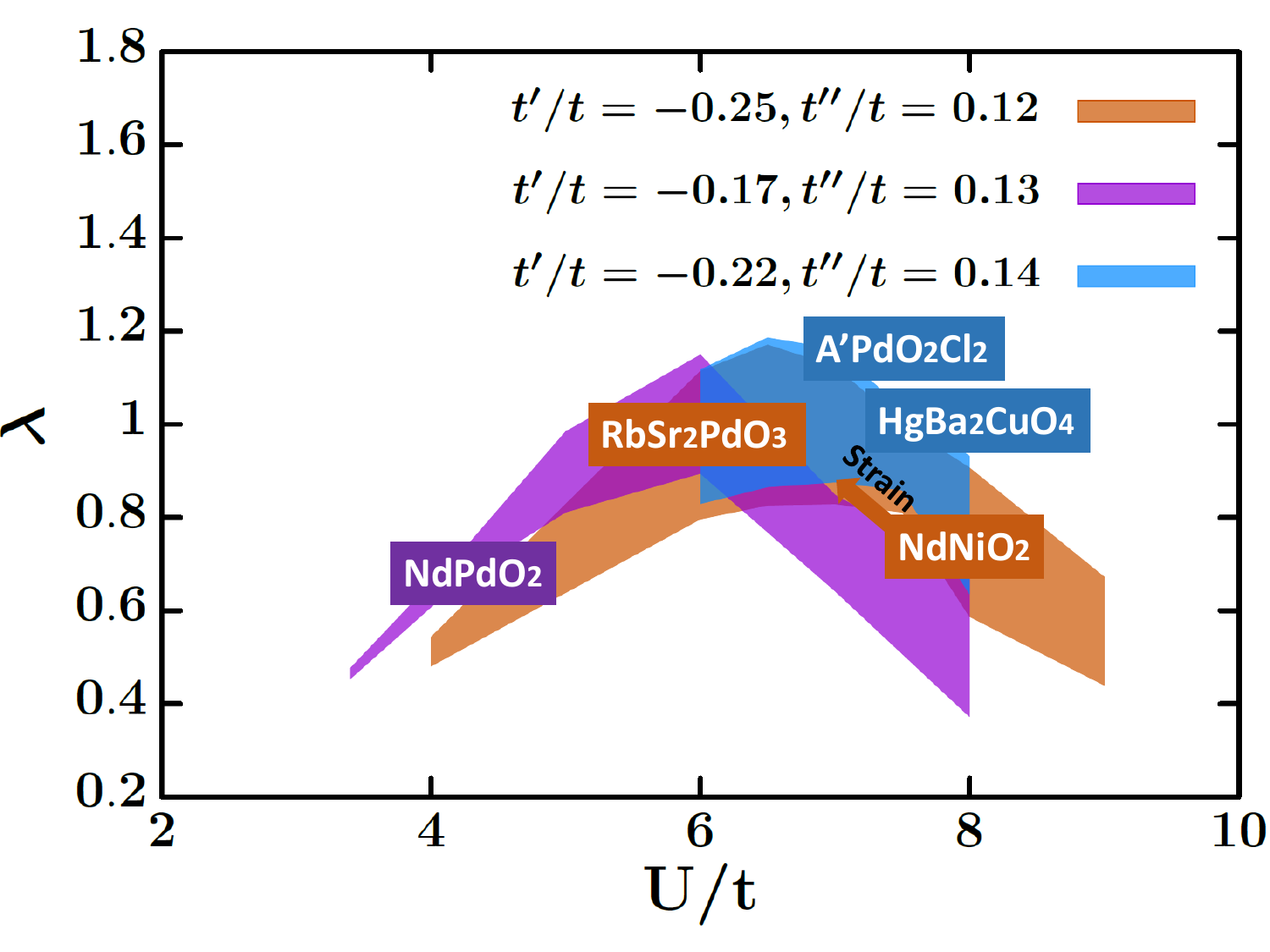}
        \caption{(Color online) Interaction  $U$ dependence of the superconducting eigenvalues $\lambda$ 
        at $T=0.01t$ for three different $t'$,$t''$ from Table~\ref{tab:my_label} . For each $U$, the linewidth corresponds to the range of $\lambda$ for fillings $0.80 \le n \le 0.90$. Materials corresponding to these models and $U$'s are indicated.
        \label{Fig:Udep}}
\end{figure}

Fig.~\ref{Fig:Udep}  demonstrates that a dome shape of $\lambda$ develops as a function of $U$. 
The optimal $U$ slightly depends on the 
$(t^{\prime},t^{\prime\prime})$ parameters. This trend can be explained by the earlier onset of the pseudogap for systems with small $t^{\prime},t^{\prime\prime}$, due to better (antiferromagentic) nesting.

In Fig.~\ref{Fig:Udep}, we also show the  points corresponding to each material from Table~\ref{tab:my_label}. Further we include HgBa$_2$CuO$_4$ as a typical cuprate which has $U/t \sim 7.5t-8t$ \cite{Hirayama2019,Nilsson2019,Teranishi2021} and almost the same $t^{\prime},t^{\prime\prime}$ as $A^\prime_2$PdO$_2$Cl$_2$. 
As mentioned in previous works \cite{Sakakibara2020,Kitatani2020}, we can see that NdNiO$_2$ has a too large interaction and thus falls outside the area with highest $T_{\rm c}$.

Following this insight, we can rationalize now the recent experimental achievements of realizing higher $T_{\rm c}$'s in NdNiO$_2$: both external pressure \cite{Wang2021} and in-plane lattice compressive strains \cite{Lee2022,ren2021superconductivity} play similar roles at reducing the in-plane lattice constant, shrinking the Ni-Ni distance and weakening the correlation strength $U/t$, while increasing the magnetic exchange $4t^2/U$ \cite{Chang_LSCO}. Simply replacing Ni-3$d$ by Pd-4$d$ yields NdPdO$_2$ which is too weakly correlated in Fig.~\ref{Fig:Udep}. 
 \kh{Let us emphasize that this concerns tetragonal NdPdO$_2$  which can be stabilized in thin films. In the bulk and for thick films, there will be a substantial orthorhombic distortion, see SM \cite{SM} Section VIII.  Due to the  tilting of the PdO$_6$ octahedra, $t$ is reduced to 304$\,$meV. This pushes orthorhombic
NdPdO$_2$ to $U/t\sim7$, i.e.,  close to the optimum in Fig.~\ref{Fig:Udep} (average $t'/t\approx \mk{-0.25}$; $t''/t\approx0$).  However this enhancement of $T_c/t$  is largely compensated by the smaller $t$, altogether yielding a  similar $T_c$.}

\kh{A} more promising solution to increase $U/t$ \kh{and thus $T_c$}  is to enlarge the lattice which is possible by inserting  spacing layers between the PdO$_2$ planes. Doing so, our DFT and cRPA calculations indeed place RbSr$_2$PdO$_3$ and $A^\prime_2$PdO$_2$Cl$_2$ close to the optimum. 

\begin{figure}[tbp]
        \centering
        \includegraphics[width=\linewidth,angle=0]{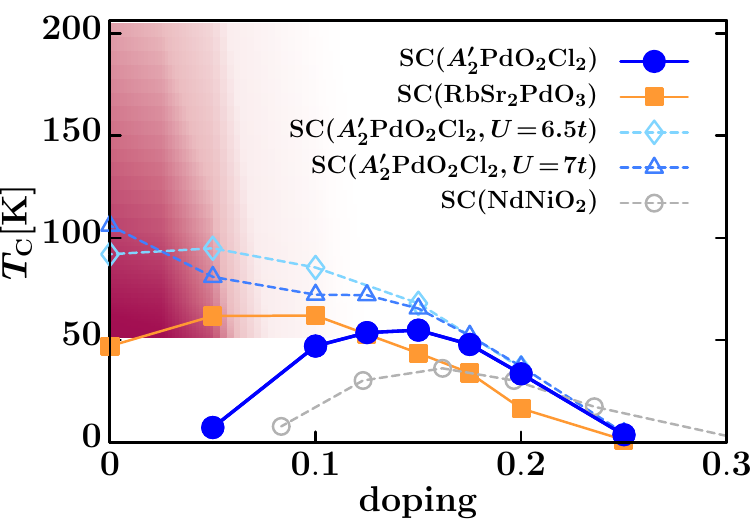}
        \caption{(Color online) Phase diagrams for nickelates (NdNiO$_2$) and palladates (RbSr$_2$PdO$_3$ and $A^{\prime}_2$PdO$_2$Cl$_2$). 
        In the red region, we expect antiferromagnetism instead of superconductivity for $A^{\prime}_2$PdO$_2$Cl$_2$ if there is a weak inter-layer coupling because of the 
        huge $\chi_{\rm sp}$ (color code \mk{of $\chi_{\rm sp}$ in D$\Gamma$A} as in Fig.~\ref{Fig:lambda}).
        \label{Fig_digram}}
\end{figure}

We finally show in Fig.~\ref{Fig_digram} the phase diagram for these palladate compounds, using the same approach as previously for nickelates \cite{Kitatani2020}. We predict palladates to have a $T_{\rm c}\gtrsim 60$\,K, which touches the floor level line of cuprates and remarkably exceeds the current upper limit of nickelates, $T_{\rm c}=30$\,K \cite{Lee2022,Wang2021}. Indeed, the calculated phase diagram (superconductivity and antiferromagnetism) for $A^{\prime}_2$PdO$_2$Cl$_2$ is quite similar to well-known cuprate phase diagrams. Furthermore,  weak strain (described by $U=6.5t,7t$ \cite{U65}, c.f., Table~\ref{tab:my_label}) would tune the material to an even higher $T_{\rm c}$.

\noindent
{\sl Conclusion and outlook}---Building on the success of  D$\Gamma$A to {\em predict} the superconducting dome in nickelates, we have performed a comprehensive survey of the Hubbard model and revealed the optimal phase-space region for unconventional superconductivity.
Conditions are optimal at intermediate coupling $(U/t=6-7)$, moderate Fermi surface warping $(|t^{\prime}|+|t^{\prime\prime}| \approx 0.3-0.4)$, and low hole doping $(n \!\sim\! 0.90-0.95)$.
Combining this insight with first principles calculations, we predict  palladates and nickelates grown on compressive substrates to be superconductors with a $T_{\rm c}$ comparable to cuprates. 

The theoretically proposed palladates have yet to be synthesized.
Palladates with a perovskite-like structure \cite{kim2002x,kim2001lapdo3}, however, have already been realized in experiment. Then, provided that a reduction process, similar to that of NdNiO$_3$ $\longrightarrow$ NdNiO$_2$, is possible, realizing palladates with PdO$_2$ layers is a promising route for high-$T_{\rm c}$ superconductors.
Additionally, the possibility to engineer cuprate analogs based on $4d$ materials has been discussed based on AgF$_2$ \cite{Gawraczynski2019} and silver oxides \cite{Hirayama2022}. Analyzing the relation between those systems and the ones we are proposing in this letter will offer fresh insight into the design of superconductors.

{\sl Acknowledgments}---We would like to thank Motoaki Hirayama and Yusuke Nomura for illuminating discussions. We thank Alaska Subedi for  providing us with lower symmetry phase structures of nickelates from Ref.~\onlinecite{Subedi2023}.
We acknowledge the financial support by Grant-in-Aids for Scientific Research (JSPS KAKENHI) Grant No. JP21K13887 and JP19H05825 as well as by projects P32044 and I5398 of the Austrian Science Funds (FWF).
Calculations have been done mainly on the Vienna Scientific Cluster (VSC).

\bibliography{full,add}

\begin{thebibliography}{139}
\bibitem [{U65()}]{U65}%
  \BibitemOpen
  \href@noop {} {}\bibinfo {note} {For drawing the $U=6.5t$ phase diagram, we
  averaged the $-1.5\%,-3.0\%$-strain result from Table
  {\ref{tab:my_label}}.}\BibitemShut {Stop}%
\end{thebibliography}
\end{document}


\title{Supplementary material for ``Optimizing superconductivity: from cuprates via nickelates to palladates"}

\author{Motoharu Kitatani$^{a,b}$, Liang Si$^{c,d}$, Paul Worm$^d$, Jan M. Tomczak$^d$, Ryotaro Arita$^{b,e}$ and Karsten Held$^d$}

\affiliation{$^a$Department of Material Science, University of Hyogo, Ako, Hyogo 678-1297, Japan}
\affiliation{$^b$RIKEN Center for Emergent Matter Sciences (CEMS), Wako, Saitama, 351-0198, Japan}
\affiliation{$^c$School of Physics, Northwest University, Xi’an 710127, China}
\affiliation{$^d$Institute of Solid State Physics, TU Wien, 1040 Vienna, Austria}
\affiliation{$^e$Research Center for Advanced Science and Technology, University of Tokyo
4-6-1, Komaba, Meguro-ku, Tokyo 153-8904, Japan}


\date{\today}

\begin{abstract}
This supplementary material contains additional results obtained by density functional theory  (DFT), dynamical mean field theory (DMFT) and dynamical vertex approximation (D$\Gamma$A) that further corroborate our conclusions of the main text and is organized in nine sections.
In Section I, we perform discussions on the crystal structures, parameters and analyze similarities and differences between the nickelate NdNiO$_2$ and the palladates NdPdO$_2$, RbSr$_2$PdO$_3$ and $A^\prime_2$PdO$_2$Cl$_2$ ($A^{\prime}={\rm La}_{0.5}{\rm Ba}_{0.5}$). Section II provides detailed DFT (and beyond DFT) results of electronic structure computations for all compounds. In Section III we show the DMFT results, including spectral functions and effective mass, for RbSr$_2$PdO$_3$ and $A^\prime_2$PdO$_2$Cl$_2$. Section IV shows the self-energy effect on the spectrum and the superconductivity instability in more detail. In Section V, we show the numerical details on the D$\Gamma$A study of superconductivity. Further, in Section VI we present our Wannier function projection  and in Section VII the  computational details of the virtual crystal approximation. Section VIII analyzes the phonon dispersion  and structural  distortions of LaPdO$_2$. Finally in Section IX we compute the  DFT total energy for different magnetic states of  $A'_2$PdO$_2$Cl$_2$.
\end{abstract}

\maketitle

\section{I. Crystal structure of nickelates and palladates}

\begin{figure}[tb]
\centering
\includegraphics[width=\linewidth,angle=0]{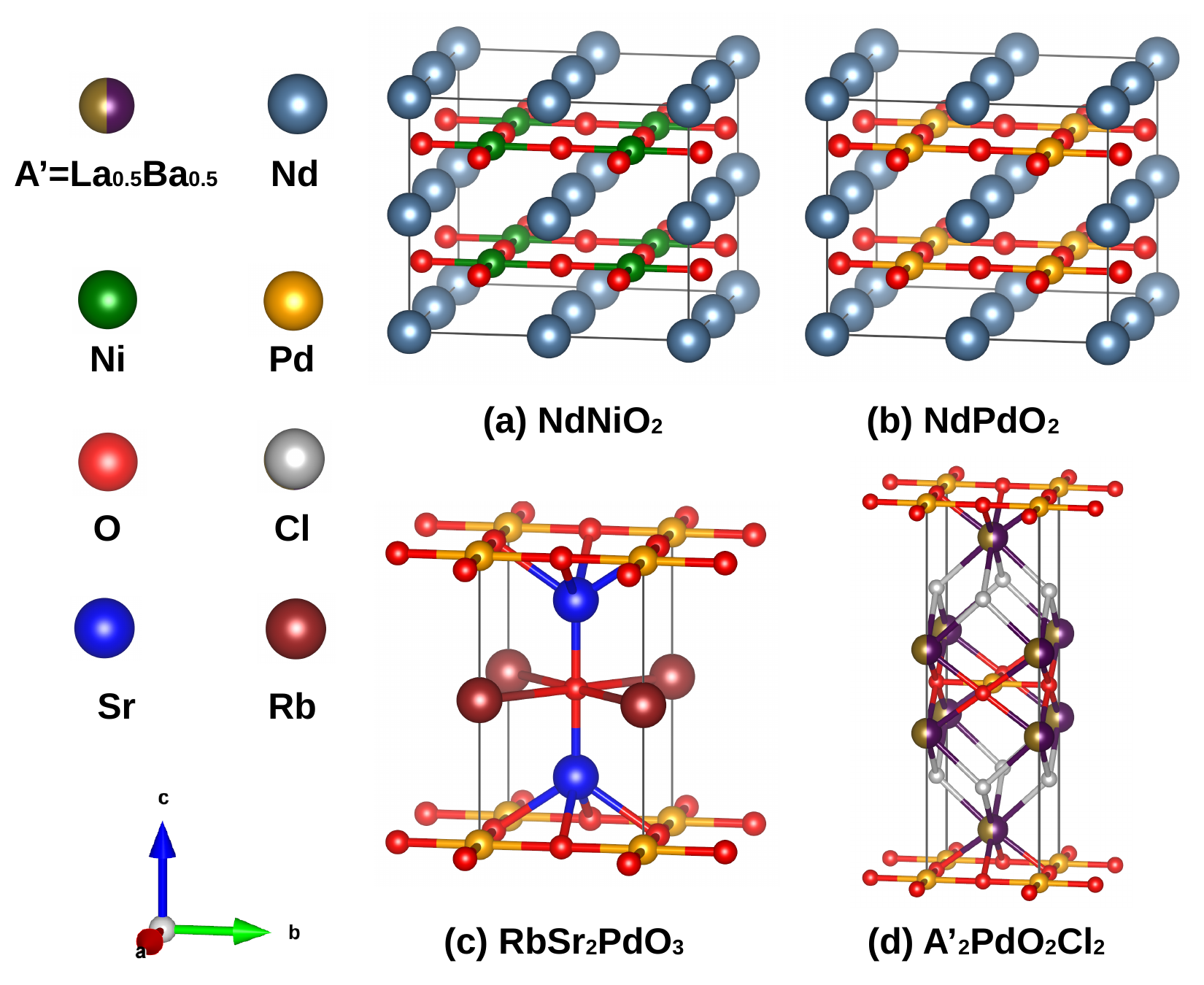}
\caption{Crystal structures of (a) NdNiO$_2$, (b) NdPdO$_2$, (c) RbSr$_2$PdO$_3$ and (d) $A^\prime_2$PdO$_2$Cl$_2$.}
\label{FigS1_structure}
\end{figure}

The crystal structures of nickelate NdNiO$_2$ and palladates NdPdO$_2$, RbSr$_2$PdO$_3$ and $A^\prime_2$PdO$_2$Cl$_2$ are shown in Fig.~\ref{FigS1_structure}. To guarantee a 3$d^9$ configuration of the Ni$^{1+}$ cation in $A^\prime_2$PdO$_2$Cl$_2$, we need a $A^\prime{}^{2.5+}$ cation which can be  achieved by mixing 50\%La and 50\%Ba. To approach optimal filling/doping in the proposed palladates  RbSr$_2$PdO$_3$ and $A^\prime_2$PdO$_2$Cl$_2$, the ration between Rb$^{1+}$/Sr$^{2+}$ for the former and La$^{3+}$/Ba$^{2+}$ for the later case can be  tuned.

The relaxed parameters for all nickelates and palldates are shown in Table~\ref{Table_lattice}. For the DFT-level calculations, PBESol \cite{PhysRevLett.100.136406} is employed for structural relaxations, and PBE \cite{PhysRevLett.77.3865} for electronic structure calculations (e.g., DFT band structures and tight-binding Wannier projections), following the lines of Ref.~\cite{Motoaki2019}. Our PBESol structural relaxations yield the lattice constants of NdNiO$_2$ $a$=$b$=3.864\,\AA~and $c$=3.243\,\AA. 
To simulate in-plane compressive strain effects, we tune the in-plane lattices to 3.800\,\AA~[of bulk LaAlO$_3$ (LAO)], the $z$-lattice is then simultaneously relaxed to 3.330\,\AA. 
assuming that the films are several layers thick as in experiments for nickelate superconductors, this change of the lattice parameters is the dominant effect of the substrate; apart from that a bulk DFT calculation is justified.
The corresponding changes of hoppings and $U$ are shown in Table~I of the main text.

For NdPdO$_2$, an in-plane lattice expansion from 3.864\,\AA~to 4.085\,\AA~ results from substituting Ni by Pd. Such an expansion may contribute to smaller orbital overlap and stronger electronic correlations if the $B$-site atoms are unchanged, however the more delocalized 4$d$-orbitals of Pd counteract this with their larger orbital spreads (see main text). When the in-plane lattices are fixed to the one of SrTiO$_3$ to simulate the strain effects from a STO substrate, the out-of-plane is  relaxed to 3.255\,\AA, which boosts its two-dimensional characters.

\begin{table*}[tb]
    \caption{DFT-PBESol relaxed lattice parameters of NdNiO$_2$, NdPdO$_2$ (with full relaxation or grown on SrTiO$_3$ substrate), RbSr$_2$PdO$_3$ and $A^\prime_2$PdO$_2$Cl$_2$ (in units of \AA). $A^{\prime}$ is La$_{0.5}$Ba$_{0.5}$. In the last column we show the appropriate substrates, in the brackets are the in-plane lattice constants (also in unit of \AA).}
    \label{Table_lattice}
    \centering
    \begin{tabular}{c|c|c|c|c|c}
        \hline 
        \hline
         - & $a$/$b$ & $c$ & Ni-Ni/Pd-Pd (in-plane) & out-of-plane & Substrate \\
        \hline
        NdNiO$_2$ & 3.864 & 3.243 & 3.864 & 3.243 & LSAT (3.868) \\
        \hline
        NdNiO$_2$ (strained: on LaAlO$_3$) &   3.800 & 3.330 & 3.800 & 3.330 & LaAlO$_3$ (3.80)  \\
        \hline
        NdPdO$_2$  & 4.085 & 3.143 &  4.085&  3.143 & PrScO$_3$ (4.02)  \\
        \hline
        NdPdO$_2$ (strained: on SrTiO$_3$) & 3.905 &  3.255 & 3.905 & 3.255 & SrTiO$_3$ (3.905)  \\
        \hline
        RbSr$_2$PdO$_3$ & 4.264 &  7.555 & 4.264 & 7.555 & MgO (4.20)\\
        \hline
        $A^\prime_2$PdO$_2$Cl$_2$ (unstrained) & 4.366 &  14.639 & 4.366 & 7.944 & Rutile-TiO$_2$ (4.59)/MgO(4.20) \\
        \hline
        $A^\prime_2$PdO$_2$Cl$_2$ (-1.5\% strain) & 4.301 &  14.740 & 4.301 & 7.973 & MgO (4.20) \\
        \hline
        $A^\prime_2$PdO$_2$Cl$_2$ (-3\% strain) & 4.235 &  14.870 & 4.235 &  8.016 &  MgO (4.20)  \\
        \hline
        \hline
    \end{tabular}
\end{table*}

For the  proposed palladates RbSr$_2$PdO$_3$ and $A^\prime_2$PdO$_2$Cl$_2$, the in-plane lattice constants are even larger than those of NdPdO$_2$. 
Another notable difference is  the distance between Ni-Ni or Pd-Pd along the $z$-direction (out-of-plane); it increases to $\sim$7.5-8.0\,\AA, as shown in Table~\ref{Table_lattice}. In the last column we list some suitable substrates for the proposed materials, e.g., for RbSr$_2$PdO$_3$ and strained (-1.5\% and 3\%) $A^\prime_2$PdO$_2$Cl$_2$ that host larger in-plane lattice constants than SrTiO$_3$ (3.905\,\AA), MgO is considerable to be employed as substrate as its in-plane lattice is $\sim$4.2\,\AA. For the fully relaxed NdPdO$_2$ (4.085\,\AA), PrScO$_3$ (in-plane lattice 4.02\,\AA) is possible candidate. For compressive cases (when compared with the in-plane lattice of SrTiO$_3$: 3.905\,\AA) such as unstrained NdNiO$_2$ and strained NdNiO$_2$, one of the possible substrates (LaAlO$_3$)$_{0.3}$(SrAl$_{0.5}$Ta$_{0.5}$O$_3$)$_{0.7}$ (LSAT: 3.868\,\AA) had been proved as helpful to enhance its $T_{\rm c}$ \cite{Lee2022}. For the strained case, LaAlO$_3$ (3.80\,\AA) is expected to be more effective at eliminating atomic defects and enhancing $T_{\rm c}$. The in-plane lattice constant of unstrained $A^\prime_2$PdO$_2$Cl$_2$ is predicted as 4.366\,\AA~in DFT structural relaxation, we hence proposed rutile-TiO$_2$ (4.59\,\AA) and MgO (4.20\,\AA) as a possible substrate in realistic experiments.

\section{II. Electronic structure of nickelates and palladates}

To investigate the difference between nickelates and palladates, we first compute their band centroids. The bands centroids in Table~\ref{Table_center} are computed by $E_i=\int g_i(E)EdE/\int g_i (E)$, here $g_i$ is the partial density of states of the corresponding orbital $i$ and $E$ is the energy. The integration ranges covers both the bonding and anti-bonding states for Cu/Ni/Pd-$d$ orbital and O-$p$ orbitals (up to 5, 7 and 7\,eV for CaCuO$_2$, NdNiO$_2$ and NdPdO$_2$, respectively). As one can see, for the cuprate CaCuO$_2$, both the centroids of Cu-$d$ and $d_{x^2-y^2}$ are just $\sim$0.4\,eV and $\sim$0.7\,eV higher than that of O-$p$. This is consistent with the fact that  CaCuO$_2$  is a charge-transfer insulator.

For NdNiO$_2$, the $d$-bands ($d_{x^2-y^2}$ band) shift up by $\sim$1\,eV ($\sim$0.9\,eV) compared with those in CaCuO$_2$. Moreover, the O-$p$ bands are shifted down by $\sim$0.9\,eV. As a consequence, the charge transfer energies $\Delta(d-p)$ and $\Delta(d_{x^2-y^2}-p)$ in NdNiO$_2$ are $\sim$1.8\,eV larger than in CaCuO$_2$.

For the palladate NdPdO$_2$, the band centroid of $d$-orbitals $E(d)$ is comparable with those of CaCuO$_2$ while  $E(d_{x^2-y^2})$ is $\sim$0.3\,eV higher than in CaCuO$_2$. Thus, the charge transfer energies $\Delta(d-p)$ and $\Delta(d_{x^2-y^2}-p)$ are 1.211 and 1.827\,eV, i.e., in between the charge transfer energies of CaCuO$_2$ and NdNiO$_2$. The computations strongly hint that NdPdO$_2$ hosts a similar 2D nature as in both CaCuO$_2$ and NdNiO$_2$.

\begin{figure*}[tb]
\centering
\includegraphics[width=\linewidth,angle=0]{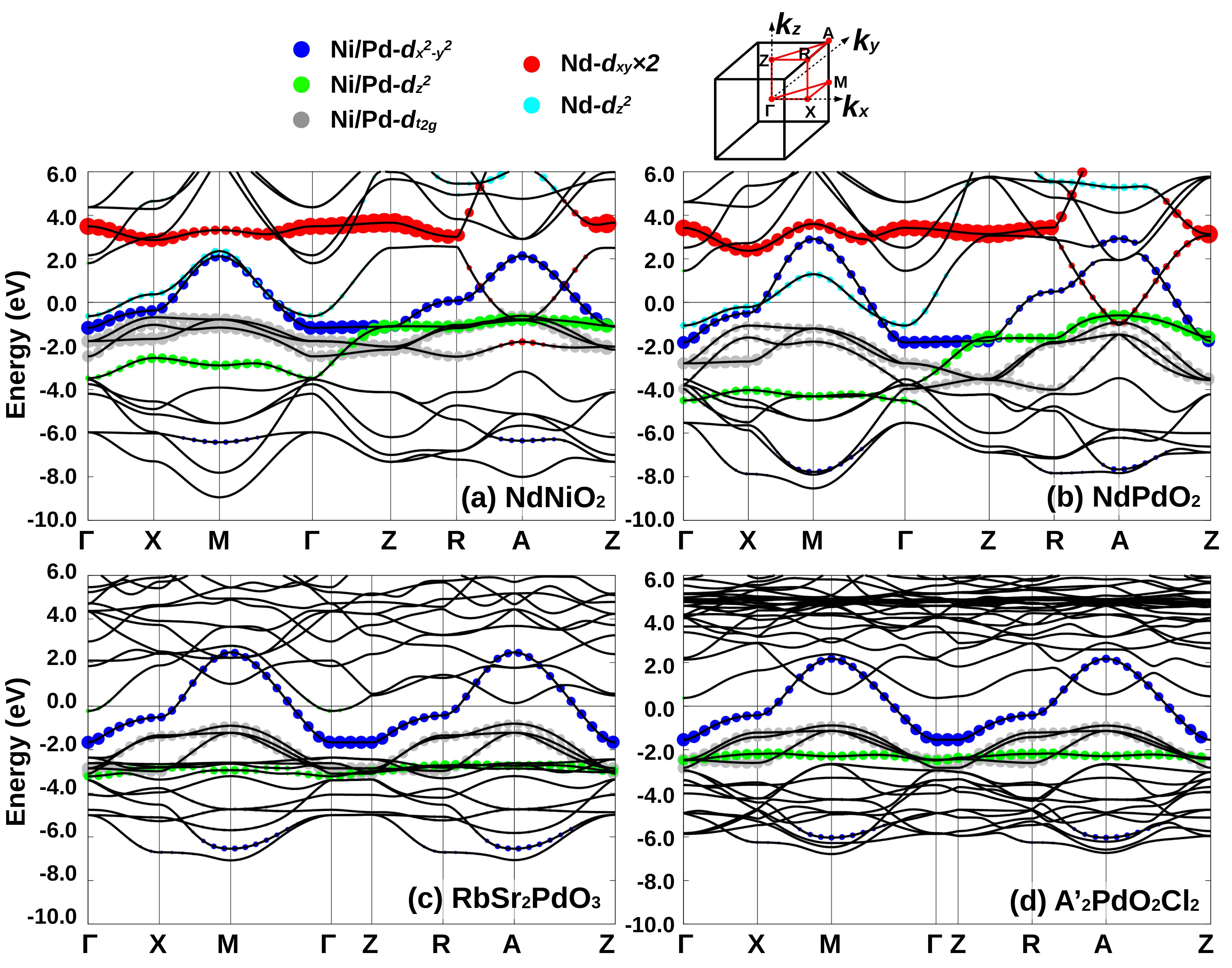}
\caption{DFT band characters of (a) NdNiO$_2$, (b) NdPdO$_2$, (c) RbSr$_2$PdO$_3$ and (d) A$^{\prime}$PdO$_2$Cl$_2$. Here, we used the fully relaxed structure for all systems. The top panel shows the symbols for different orbitals and the first Brillouin zone. The size of the symbols for Nd-$d_{xy}$ are renormalized by a factor of 2 for clearness.}
\label{FigS2}
\end{figure*}

\begin{figure*}[tb]
\centering
\includegraphics[width=\linewidth,angle=0]{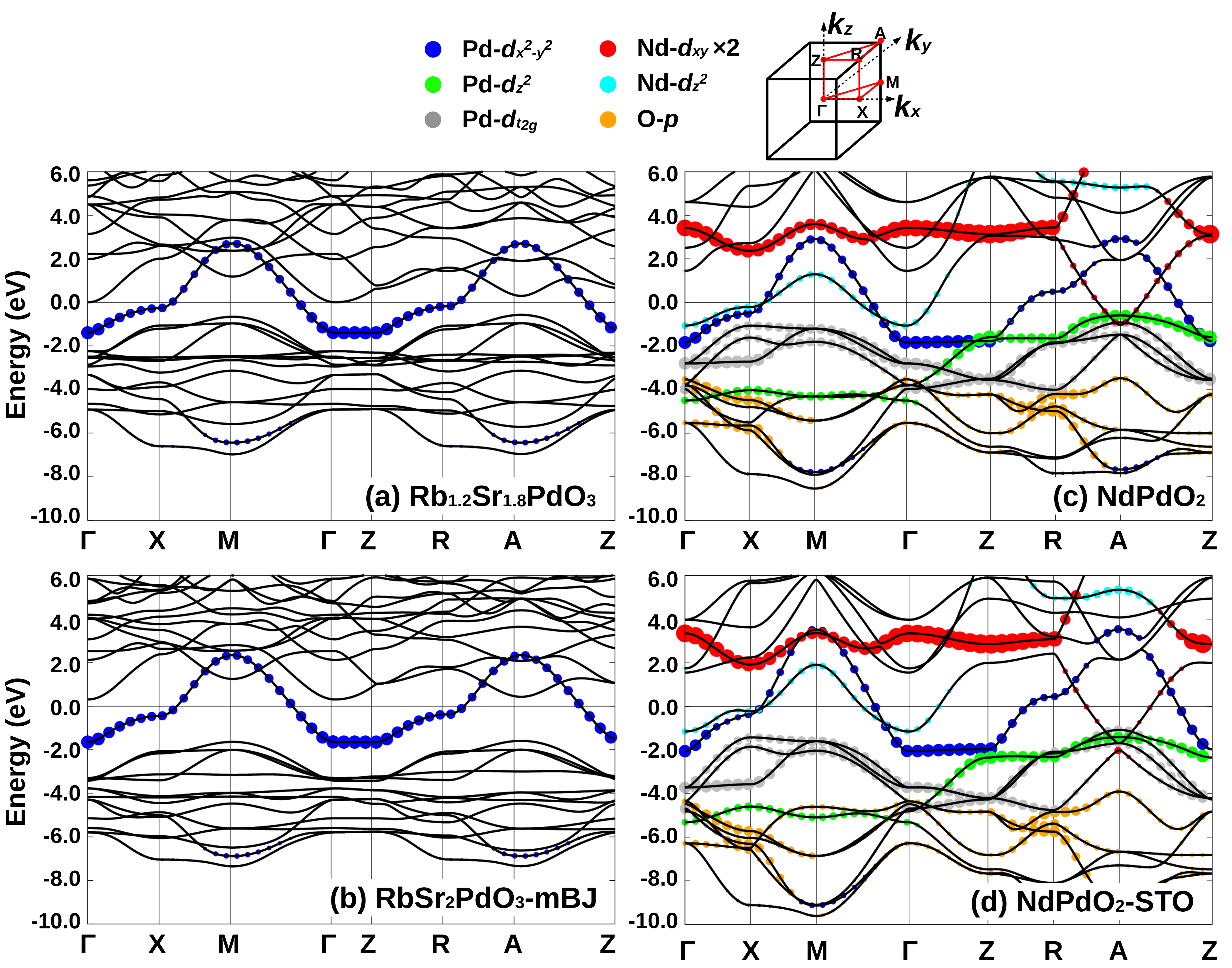}
\caption{Comparison between the DFT band characters of (a) hole doped Rb$_{1.2 }$Sr$_1.8$PdO$_3$, (b) RbSr$_2$PdO$_3$ with mBJ exchange-correlation fucntional, (c) fully relaxed NdPdO$_2$ and (d)  NdPdO$_2$ on SrTiO$_3$ (STO) substrate. The top panel shows the symbols for contribution of the different orbitals and the first Brillouin zone. The size of the symbols for Nd-$d_{xy}$ are renormalized by a factor of 2 for clearness.}
\label{FigS3_STO}
\end{figure*}

In Fig.~\ref{FigS2} we show the DFT-level band structures of NdPdO$_2$ and RbSr$_2$PdO$_3$ and $A^\prime_2$PdO$_2$Cl$_2$. NdPdO$_2$ has a larger $d_{x^2-y^2}$ bandwidth than NdNiO$_2$, which is due to its more delocalized 4$d$ orbitals nature. Its pockets at $\Gamma$ and $A$ are energetically deeper than in NdNiO$_2$. However, as concluded by previous research \cite{Kitatani2020}, these pockets are not essential for emergent nickelate superconductivity. Additionally, the band structures of $A^\prime_2$PdO$_2$Cl$_2$ [Fig.~\ref{FigS2}(d)] exhibits a Fermi structure with only the  Pd-$d_{x^2-y^2}$ orbital. For RbSr$_2$PdO$_3$ [Fig.~\ref{FigS2}(c)], the pockets merely touch the Fermi energy at the $A$ and $\Gamma$ momentum.
Indeed, the tiny $\Gamma$ pocket for RbSr$_2$PdO$_3$  is expected to become unoccupied upon hole doping or with exchange-correlations by using functional beyond standard DFT-PBE: the $\Gamma$-pocket is absent, e.g., for Rb$_{1.2}$Sr$_{1.8}$PdO$_3$ [Fig.~\ref{FigS3_STO}(a)] and RbSr$_{2}$PdO$_3$ computed by employing the improved version of the modified Becke-Johnson (mBJ) exchange potential [Fig.~\ref{FigS3_STO}(b)]. On a technical note, the process of hole doping and consequently the noninteger filling at Pd-$d_{x^2-y^2}$ in RbSr$_2$PdO$_3$ is achieved here by employing virtual crystal approximation \cite{PhysRevB.61.7877}. 

In Fig.~\ref{FigS3_STO},
the band structures of NdPdO$_2$ [Fig.~\ref{FigS3_STO}(d)] on a SrTiO$_3$ substrate, RbSr$_2$PdO$_3$ upon 0.2 hole doping per Pb (i.e., Rb$_{1.2}$Sr$_{1.8}$PdO$_3$) [Fig.~\ref{FigS3_STO}(a)], and the bands calculated by improved version of the modified Becke-Johnson exchange potential \cite{PhysRevLett.102.226401} [Fig.~\ref{FigS3_STO}(b)] are shown; for a comparison further the band of fully relaxed NdPdO$_2$ are presented in Fig.~\ref{FigS3_STO}(c). As shown in Fig.~\ref{FigS3_STO}(a-b), both hole doping and exchange-correlation potentials are shift the $\Gamma$ pocket of RbSr$_2$PdO$_3$ above the Fermi energy E$_f$. These  results demonstrate that the realistic Fermi surface of RbSr$_{2}$PdO$_3$ can also be expected to be composed of a single band with $d_{x^2-y^2}$ character, even without additional hole doping. The correlations from Pd-4$d$ and the exchange effects, which are both only rudimentarily accounted for in DFT, play an effective role at preserving a 2D cuprates-like Fermi surface.

As we discussed in the main text, the fully relaxed in-plane lattice constants of NdPdO$_2$ are $a$=$b$=4.085\,\AA, 4.6\% more than a  SrTiO$_3$ substrate: 3.905\,\AA. A suitable substrate for an almost unstrained film is PrScO$_3$ whose lattice is 4.02\,\AA~(as shown in Table~\ref{Table_lattice}). If NdPdO$_2$ is grown on SrTiO$_3$, the in-plane lattice is fixed as 3.905\,\AA. As a consequence, the out-of-plane lattice is relaxed to larger value of 3.255\,\AA,  and thus a better 2D character of the  $d_{x^2-y^2}$ orbital. As shown in [Fig.~\ref{FigS3_STO}(c-d)], such a strain does not change the main character of the Pd-$d_{x^2-y^2}$ orbital.

\begin{table}[tb]
    \caption{DFT calculated band centeriods (in  units of eV) of CaCuO$_2$, NdNiO$_2$ and NdPdO$_2$ and charge transfer energies $\Delta$.}
    \label{Table_center}
    \centering
    \begin{tabular}{c|c|c|c|c|c}
        \hline 
        \hline
         - & $E(d)$ & $E(d_{x^2-y^2})$ & $E(O-p)$ & $\Delta(d-p)$ & $\Delta(d_{x^2-y^2}-p)$ \\
        \hline
        CaCuO$_2$ & -2.367 &  -2.047 & -2.772 & 0.405 & 0.725 \\
        \hline
        NdNiO$_2$ & -1.384 & -1.148 & -3.674 & 2.289 &  2.525  \\
        \hline
        NdPdO$_2$ & -2.358 &  -1.742 & -3.570 & 1.211 & 1.827 \\
        \hline
        \hline
    \end{tabular}
\end{table}

\begin{figure*}[tb]
\centering
\includegraphics[width=\linewidth,angle=0]{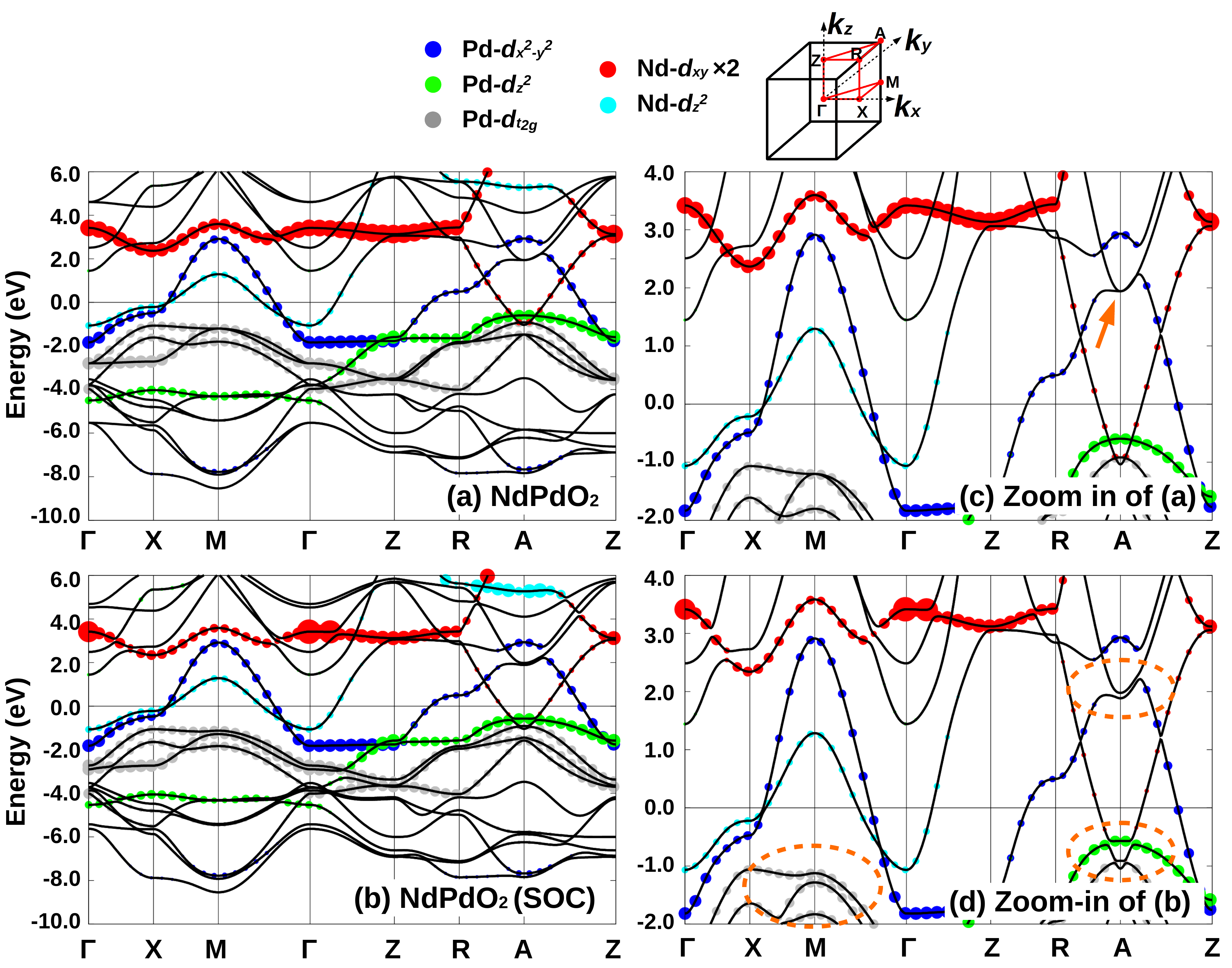}
\caption{Comparison between the DFT band characters of NdPdO$_2$ with (a,c) and without (b,d) SOC. The arrow in (c) indicates the double degeneracy of the Nd-$d_{yz}$ and Nd-$d_{xz}$ orbital at the $A$ point. The dashed orange regions in (d) indicate the SOC induced hybridization gaps.}
\label{FigS4_SOC}
\end{figure*}

Finally, as Pd is heavier than Ni, one may consider the spin-orbital coupling (SOC) effects in palladates. We hence perform DFT+SOC band computations for NdPdO$_2$ (fully relaxed case) with and without SOC, as shown in Fig.~\ref{FigS4_SOC}. Fig.~\ref{FigS4_SOC}(a) and (b) show the DFT band of NdPdO$_2$ without and with SOC, respectively. For better visibility, we also plot these bands in a smaller energy region from -2 to 4\,eV in  Fig.~\ref{FigS4_SOC}~(c,d). We find three band-crossing regions in which hybridization gaps open after switching-on SOC effect, see \ref{FigS4_SOC}(c). The first region is at $E\sim$-1 to -2\,eV and $k$=$X$-$M$. Here, the SOC opens a hybridization gap between the Pd-$t_{2g}$ bands. The second region is at $E\sim$-0.5\,eV and $k$=$A$, where there is a band-crossing between Nd-$d_{xy}$ orbitals, which mostly contribute to the $A$-pocket, and Pd-$t_{2g}$ bands. Since these two regions are irrelevant to our target Pd-$d_{x^2-y^2}$ band, the effect on superconductivity will be minor.
The third region is at $E\sim$2\,eV and $k$=$A$. In Fig.~\ref{FigS4_SOC}(c), as one see that there is a double-band degeneracy composed of Nd-$d_{yz}$ and $d_{xz}$ bands, labeled by the arrow in (c). This double degeneracy exchanges its band characters with Pd-$d_{x^2-y^2}$ band at $A$-point. And the gap opens after including SOC effect because the double degeneracy between Nd-$d_{yz}$ and Nd-$d_{xz}$ is lost, as shown in the orange region in Fig.~\ref{FigS4_SOC}(d). However, we can exclude this small band-openening as a  possible factor suppressing superconductivity in palladates because the energy of this gap is high ($\sim$2\,eV) compared to E$_f$ at the very top of the energy region of Pd-$d_{x^2-y^2}$ band (-2 to 3\,eV). Additionally, hole-doping will further reduce E$_f$ and weaken this hybridization.

All in all, we conclude that for the low energy physics around the Fermi level, the SOC does not play an important role. This is because there is only a single  Pd-$d_{x^2-y^2}$ band crossing the Fermi level. While SOC could be disadvantageous for superconductivity, in principle, for superconductivity in the studied palladates it does not seem to be relevant. Hence, we do not consider it in the subsequent DMFT and D$\Gamma$A calculations.

\section{III. Full $d$-shell projection in DMFT}
The DFT bands are projected onto maximumly localized Wannier functions  \cite{PhysRev.52.191,PhysRevB.56.12847,PhysRevB.65.035109,RevModPhys.84.1419} of (i) only 
the Pd $d_{x^2-y^2}$ band (for the single-band DMFT and D$\Gamma$A calculations presented in the main text) and of (ii) all Pd $d$ bands (for the full-$d$ set calculations in this section),  using the  \textsc{Wannier90} \cite{mostofi2008wannier90} and \textsc{Wien2Wannier} \cite{kunevs2010wien2wannier} codes. These corresponding low-energy effective tight-binding Hamiltonians are generated and subsequently supplemented by a local Kanamori interaction, using the fully localized limit as double counting \cite{PhysRevB.48.16929}. DMFT calculations in the present research are carried out  at 300\,K with the \textsc{W2dynamics} code \cite{wallerberger2019w2dynamics}, which solves the corresponding impurity problem using the continuous time quantum Monte Carlo (CTQMC) approach in the hybridization expansion (CT-HYB) \cite{RevModPhys.83.349}. Real-frequency spectra are obtained with the \textsc{ana\_cont} code \cite{Kaufmann2021} via analytic continuation using the maximum entropy method (MaxEnt) \cite{PhysRevB.44.6011,PhysRevB.57.10287}.

\begin{figure*}[tb!]
\centering
\includegraphics[width=\linewidth,angle=0]{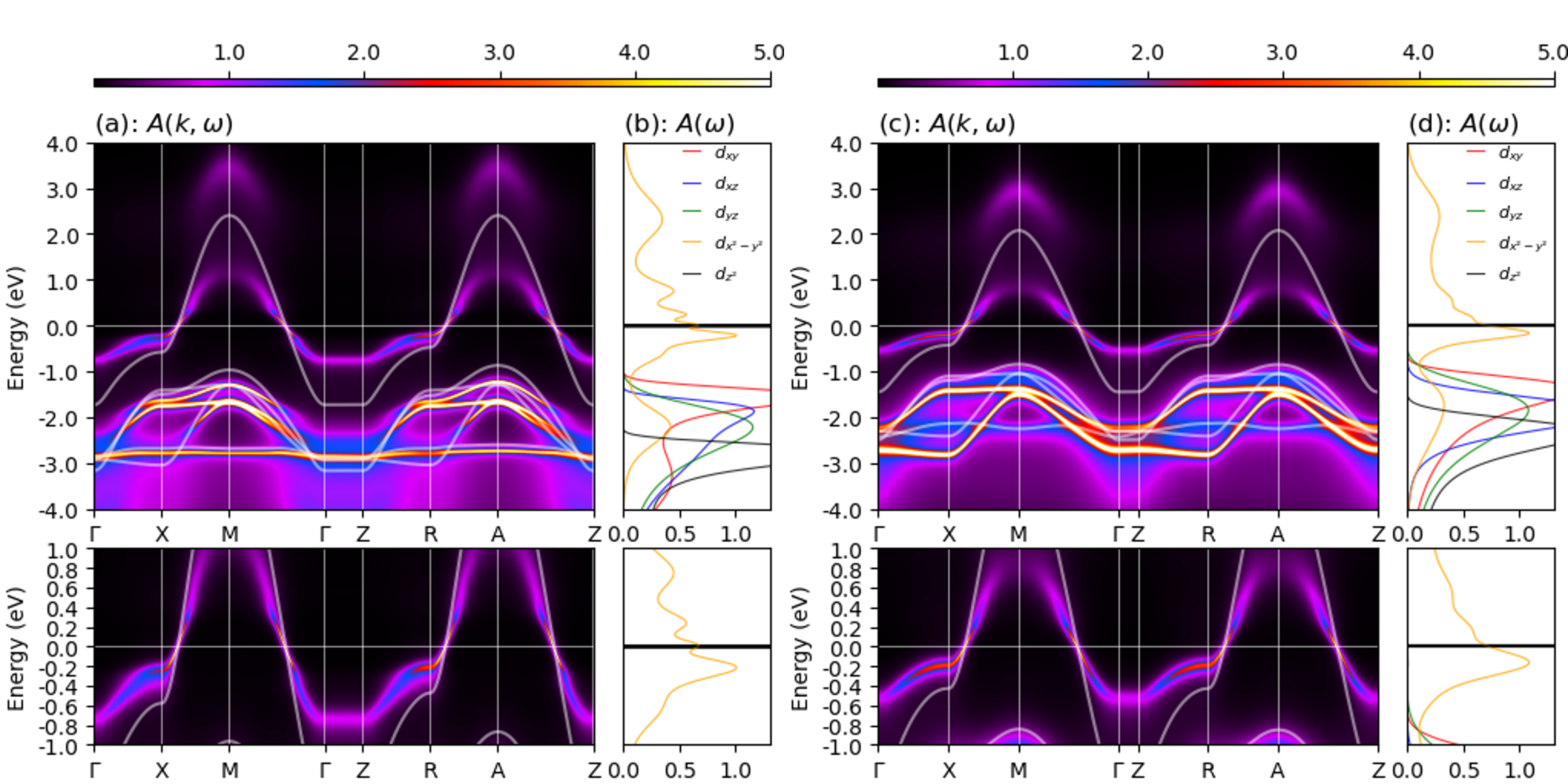}
\caption{Spectral function, $k$-resolved ($A(k,\omega)$) and $k$-integrated  ($A(\omega)$), for a full $d$-shell projection at a nominal filling of $n=9.0$. 
(Left) RdSr$_2$PdO$_3$ and (right) $A^\prime_2$PdO$_2$Cl$_2$.}
\label{Fig:FullShellDMFT-1}
\end{figure*}
\begin{figure*}[tb!]
\centering
\includegraphics[width=\linewidth,angle=0]{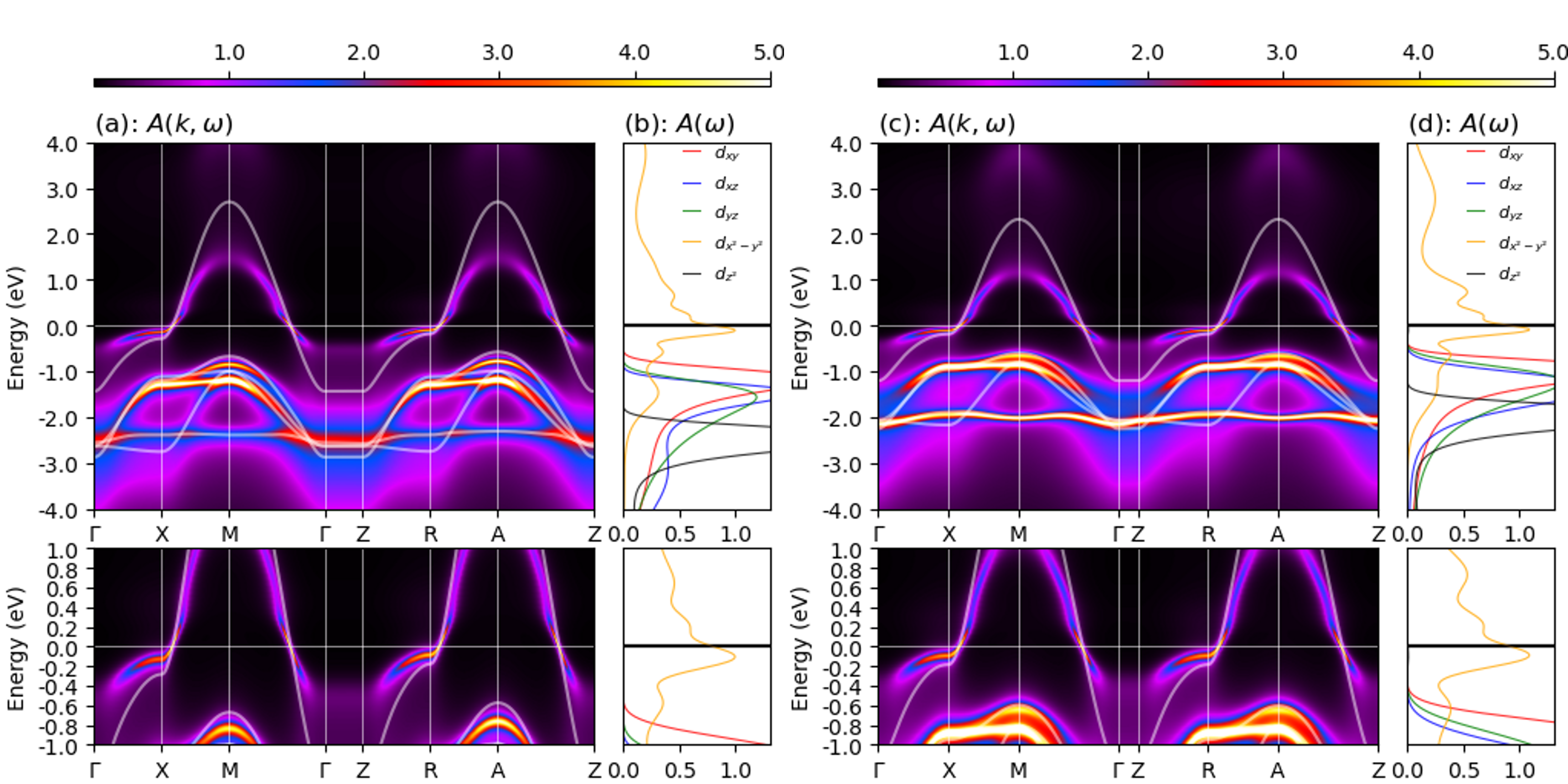}
\caption{Same as Fig.~\ref{Fig:FullShellDMFT-1} but now at a nominal filling of $n=8.8$, which corresponds to the superconducting region.}
\label{Fig:FullShellDMFT-2}
\end{figure*}

\begin{figure}[tb!]
\centering
\includegraphics[width=\linewidth,angle=0]{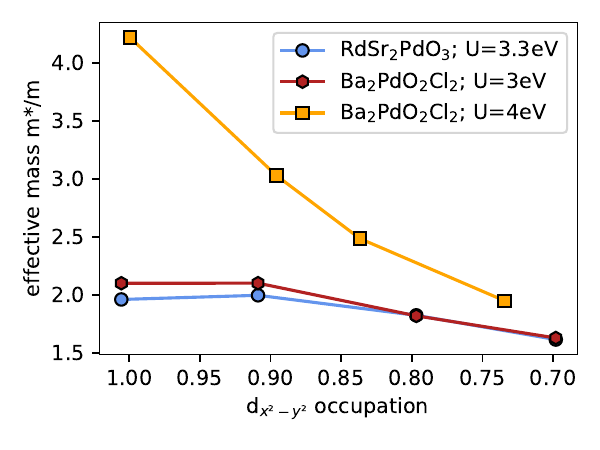}
\caption{Effective mass (m$^*$/m) as a function of d$_{x^2-y^2}$ occupation for RdSr$_2$PdO$_3$ at $U=3.3$eV (blue) and Ba$_2$PdO$_2$Cl$_2$ at  $U=3$eV (red) and $U=4$eV (orange).}
\label{Fig:EffectiveMass}
\end{figure}

The corresponding DMFT results of RdSr$_2$PdO$_3$ and $A^\prime_2$PdO$_2$Cl$_2$ are shown in Fig.~\ref{Fig:FullShellDMFT-1}, Fig.~\ref{Fig:FullShellDMFT-2} and Fig.~\ref{Fig:EffectiveMass}. Fig.~\ref{Fig:FullShellDMFT-1} shows the DMFT  $k$-resolved ($A(k,\omega)$) and $k$-integrated spectral function ($A(\omega)$) spectral functions of
the parent compounds RdSr$_2$PdO$_3$ [Fig.~\ref{Fig:FullShellDMFT-1}(a)] and $A^\prime_2$PdO$_2$Cl$_2$ [Fig.~\ref{Fig:FullShellDMFT-1}(b)], i.e.,  at 4$d^9$ Pd electronic configuration which corresponds to a half-filled Pd-$d_{x^2-y^2}$ orbital. Inclduing DMFT electronic   correlations, the other $d$-bands ($t_{2g}$+$d_{z^2}$) of Pd are well separated from the $d_{x^2-y^2}$ band. This makes both RdSr$_2$PdO$_3$ and $A^\prime_2$PdO$_2$Cl$_2$
electronically similar to cuprates superconductors.

Large hole doping had been proved being able to destroy the single-band picture and suppress superconductivity states in both cuprates and nickelates. Hence it is worth to investigate the electronic structures of RdSr$_2$PdO$_3$ and $A^\prime_2$PdO$_2$Cl$_2$ with a certain amount of hole doping, in particular, in the hole doping region were supercodnuctivity may be expected. Here, we show thee DMFT spectra of Rd$_{1.2}$Sr$_{1.8}$PdO$_3$ (4$d^{8.8}$) and $A^\prime_2$PdO$_2$Cl$_2$ (with ${A^\prime}^{2.4+}$=(La$_{0.4}$Ba$_{0.6}$)$^{2.4+}$ and consequently 4$d^{8.8}$).  For both compounds, the single-band $d_{x^2-y^2}$ Fermi surface is conserved. 

Finally, we investigate the effective mass (m$^*$) of both compounds. Fig.~\ref{Fig:EffectiveMass} displays the mass renormalization of the $d_{x^2-y^2}$ band  as obtained in DMFT (including all Pd $d$ orbitals). For the $A^\prime_2$PdO$_2$Cl$_2$ compound we show two values of the interaction strength: $U$=3.3\,eV for RdSr$_2$PdO$_3$, 3.3\,eV and 4\,eV for$A^\prime_2$PdO$_2$Cl$_2$. As antiferromagnetic correlations become large towards half-filling so does the effective mass, in particular for the larger interaction. 
This indicates the strong tendency toward an anti-ferromagnetic Mott insulator at strong coupling. The mass renormalization for both RdSr$_2$PdO$_3$ and $A^\prime_2$PdO$_2$Cl$_2$ at optimal filling for superconductivity, i.e. $n\sim 1-0.8$, is around $m^*/m \sim 2$, which is consistent with the values of NdNiO$_2$ as $n$ approaches $\sim \! 0.82$ per Ni-$d_{x^2-y^2}$: $m^*/m=2.81$. This comparison indicates that these palladates are quite strongly correlated but less so than NdNiO2. This is advantageous for superconductivity, as discussed in the main text.

\section{IV. Effect of Fermi surface shape on $T_c$}
As shown in the main text, the self-energy weakens the spectral intensity and flattens the Fermi surface shape. Here, we discuss these effects on the Fermi surface structure in more detail, by artificially separating the different effects. 

\begin{figure*}[tbp]
        \centering
                \includegraphics[width=0.8\linewidth,angle=0]{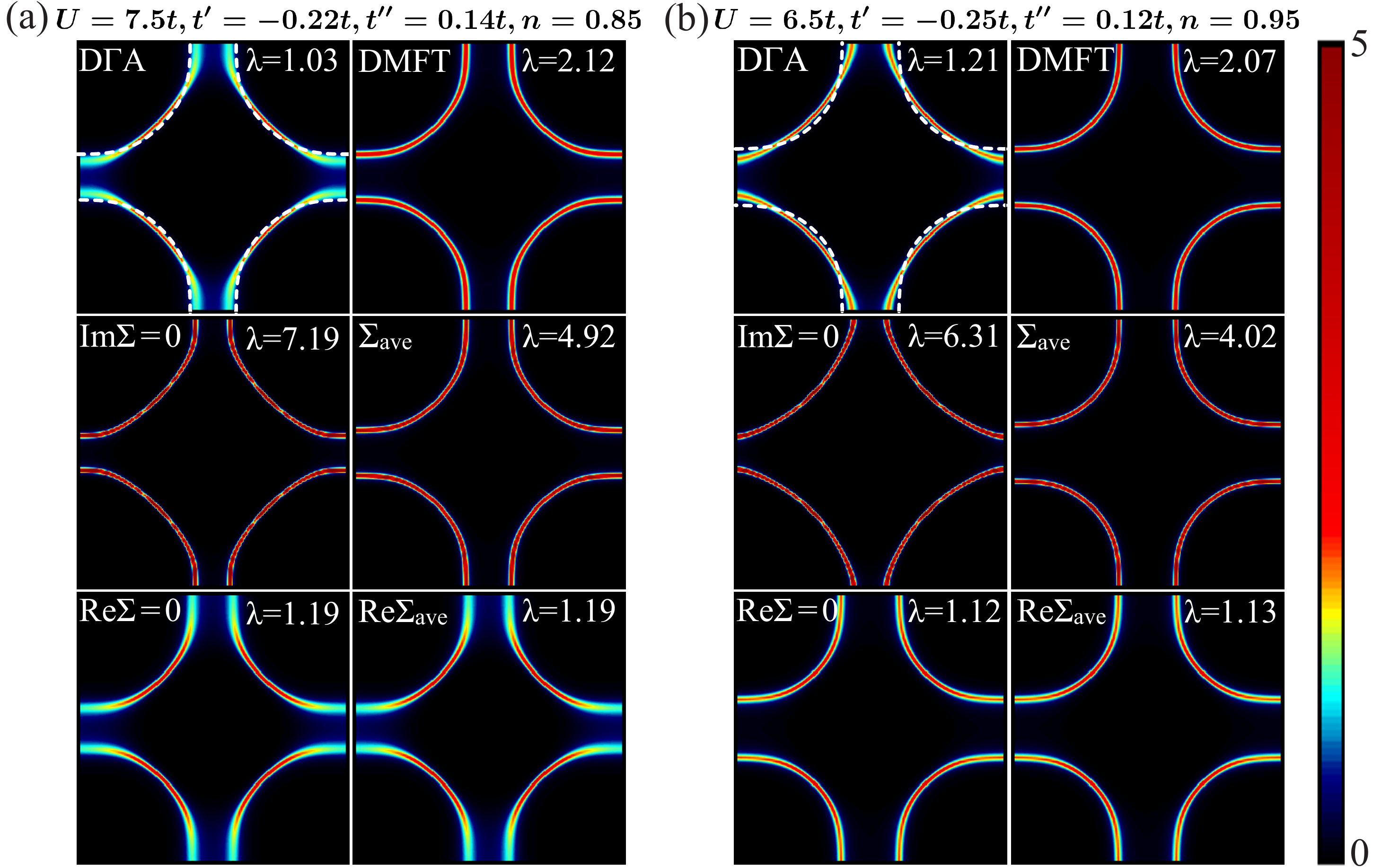}        \caption{Spectrum and superconductivity eigenvalues with DMFT, D$\Gamma$A and artificially modified self-energies (see text) at $\beta t=100$ for $U=7.5t, t^{\prime}=-0.22t, t^{\prime\prime}=0.14t, n=0.85$ (left) and $U=6.5t, t^{\prime}=-0.25t, t^{\prime\prime}=0.12t, n=0.95$ (right).}
        \label{Fig:spectrum_modifiedsigma}
\end{figure*}

In Fig.~\ref{Fig:spectrum_modifiedsigma}, we show the spectrum at two interaction strength [(a) $U=7.5t$ and (b) $U=6.5t$] comparing the full D$\Gamma$A   and DMFT self-energy
[top left (D$\Gamma$A) and top right subpanels (DMFT) in (a) and (b)] to the following  artificially modified D$\Gamma$A  self-energies: only the real part [Im$\Sigma=0$; middle left subpanels], only the imaginary part [Re$\Sigma=0$; bottom left subpanels], momentum averaged D$\Gamma$A self-energy [$\Sigma_{\rm ave}$, middle right subpanels]  and 
momentum average for the real part only, keeping the momentum-dependence of the imaginary part [Re$\Sigma_{\rm ave}$, bottom right subpanels]. We fix the filling to the original value (i.e., recalculate the chemical potential) for all cases. We also show the corresponding superconductivity eigenvalues $\lambda$ calculated with these spectra, while keeping the pairing vertex the same for solving the linearized gap equation.

We can first see that the imaginary part of the self-energy has the most relevant effect; it leads to the  pseudogap physics, as expected: From D$\Gamma$A to Im$\Sigma=0$ in Fig.~\ref{Fig:spectrum_modifiedsigma}, the shape of the Fermi surface remains the same,
but its intensity becomes stronger, in particular in the antinodal region around $(\pi,0)$. The corresponding superconducting eigenvalues is a factor 6-7 higher. 
This clearly shows that the pseudogap is counterproductive for superconductivity.

Comparing D$\Gamma$A to Im$\Sigma=0$ and the non-interacting Fermi surface (white dots in the top left subpanels) further shows that the flattening of the Fermi surface  is still captured for Im$\Sigma=0$. That is it is caused by the real part of the D$\Gamma$A self-energy,  hence there is also no flattening in the Re$\Sigma=0$ Fermi surface.

Next, if we compare the DMFT self-energy and the momentum averaged D$\Gamma$A self-energy, they give similar results. That is, the Fermi surface structure is similar to the non-interaction line (white dots in
top left subpanels). Some self-energy damping exists in both cases, but as a matter of course, there is no momentum differentiation and hence no pseudogap. In this case, the superconductivity eigenvalues are in between those for Im$\Sigma=0$ and the D$\Gamma$A self-energy. 

When ignoring the momentum dependence of the real part of the self-energy, Re$\Sigma=0$ and Re$\Sigma_{\rm k-ave}$ give similar results in  Fig.~\ref{Fig:spectrum_modifiedsigma}. The Fermi surface is the same as in the non-interacting case, but strong damping exists around the anti-nodal $(\pi,0)$ region (and symmetrically related momenta). This is a consequence of the momentum differentiation of the imaginary part of the D$\Gamma$A self-energy.

When we focus on the effect of the Fermi surface flattening on superconductivity, we can  compare the superconducting eigenvalue $\lambda$ for Re$\Sigma=0$ and Re$\Sigma_{\rm k-ave}$ with the original D$\Gamma$A $\lambda$. We surprisingly find the flattening of the Fermi surface suppresses superconductivity at $U=7.5t,t^{\prime}=-0.22t,t^{\prime\prime}=0.14t,n=0.85$ but enhances it at $U=6.5t,t^{\prime}=-0.25t,t^{\prime\prime}=0.12t,n=0.95$. 

\begin{figure}[tbp]
        \centering
                \includegraphics[width=\linewidth,angle=0]{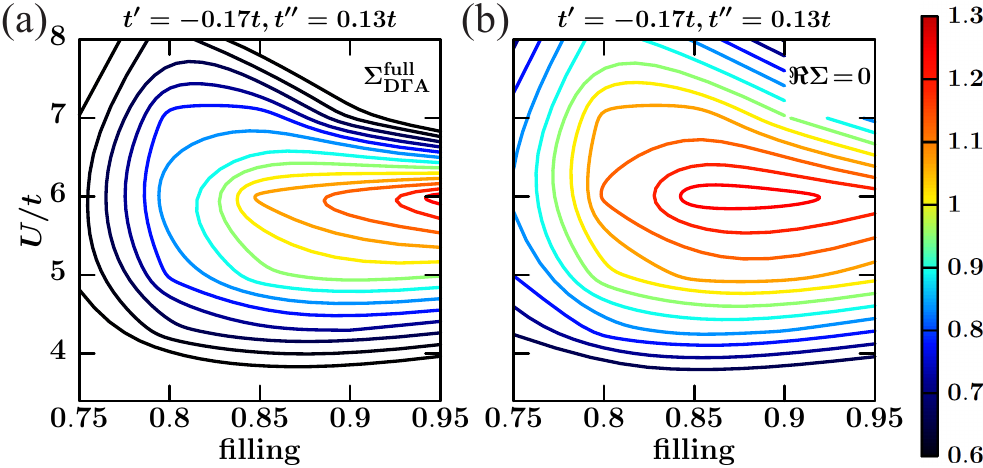}
        \caption{Comparison of superconductivity eigenvalues when using the full D$\Gamma$A self-energy ($\Sigma_{{\rm D}\Gamma{\rm A}}^{\rm full}$) and when ignoring the real-part of the self-energy ($\Re \Sigma=0$), which is responsible for flattening the Fermi surface.}
        \label{Fig:eigenvalue_modifiedsigma}
\end{figure}

To study this dichotomy further, We  analyze the parameter dependence in Fig.~\ref{Fig:eigenvalue_modifiedsigma}. We can see that there is almost no effect of Re$\Sigma$ in the weak coupling regime. In contrast, at strong coupling, the eigenvalues are almost everywhere suppressed if the flattening of the Fermi surface is properly included in  Fig.~\ref{Fig:eigenvalue_modifiedsigma} (a).
This Fermi surface flattening, given by  the real part of the D$\Gamma$A self-energy,  enhances $T_{\rm c}$ only around the optimum in the original phase diagram in Fig.~\ref{Fig:eigenvalue_modifiedsigma}(a). That is, for intermediate $U$ and fillings closer to half filling.

This result can be understood as follows: If the strength of the spectrum is similar, i.e., if there is no pseudogap, spectral weight closer to $(\pi,0)$ is favorable because the ($d$-wave) gap function has a peak there and because the van Hove singularity results in a particular strong spectral contribution.  On the other hand, if the pseudogap opens around  $(\pi,0)$, this trend will eventually reverse. For weak coupling, the Fermi surface flattening does not occur nor does the pseudogap open. There is hence no effect. For larger interactions, the flattening of the Fermi surface moves the Fermi surface toward $(\pi,0)$.
At intermediate coupling, there is no pseudogap yet 
[see Fig.~\ref{Fig:spectrum_modifiedsigma} (b)]
and superconductivity becomes enhanced. This leads to
the optimum (or hot spot) for superconductivity in
Fig.~\ref{Fig:eigenvalue_modifiedsigma} (a)
at $U\approx 6t$ and $n\approx0.95$. 
In contrast for larger interactions a pseudogap opens
[see Fig.~\ref{Fig:spectrum_modifiedsigma} (a)].
Thus despite the flattened Fermi surface, superconductivity is suppressed. If it was possible to
switch of this peudogap [as in the artificial 
Im$\Sigma=0$ panel of Fig.~\ref{Fig:spectrum_modifiedsigma} (a)]
we would get the strongest tendency toward superconductivity in this parameter regime.

\section{V. Details of D$\Gamma$A calculation}
    In this section, we explain computational details for D$\Gamma$A results. For further details and the general procedure how to calculate spin (and charge)  fluctuations and from these superconductivity in ($\lambda$-corrected), we refer the reader to \cite{Kitatani2020,Kitatani2022}.
    
    First, for solving the impurity problem and calculating local vertices, which act as a staring point, we use the continuous-time quantum Monte-Carlo (CT-QMC) on w2dynamics \cite{wallerberger2019w2dynamics} as an impurity solver [except for $U=9t$ results, which are taken from the previous results \cite{Kitatani2020} using exact diagonalization (ED)]. We have also checked that both solvers give quantitatively consistent results at $U=8t,t^{\prime}=-0.25,t^{\prime\prime}=0.12$.

In the D$\Gamma$A calculations, we separate the Matsubara frequency range into two parts following the previous works \cite{Kitatani2020,Kitatani2022}. We use 120 (80) Matsubara grids for the case $\beta t=100, U \ge 5t$ (others) for which the vertices are treated  directly. For an extended grid of 2048 Matsubara frequencies on  the positive side we use the bare-$U$ contribution instead of the local vertex. This is sufficient to obtain well-converged results even at low temperatures \cite{Kitatani2019}. For momentum grids, we take a $120 \times 120$ momentum mesh, which may not be enough to get quantitatively converged results in the critically fluctuating regime (dark magenta region in Fig.3), but elsewhere is sufficient \cite{Schaefer2019}. Nevertheless, this mesh is enough to study the superconductivity dominating region, which is the main scope of this paper.

\begin{figure}[tbp]
        \centering
                \includegraphics[width=\linewidth,angle=0]{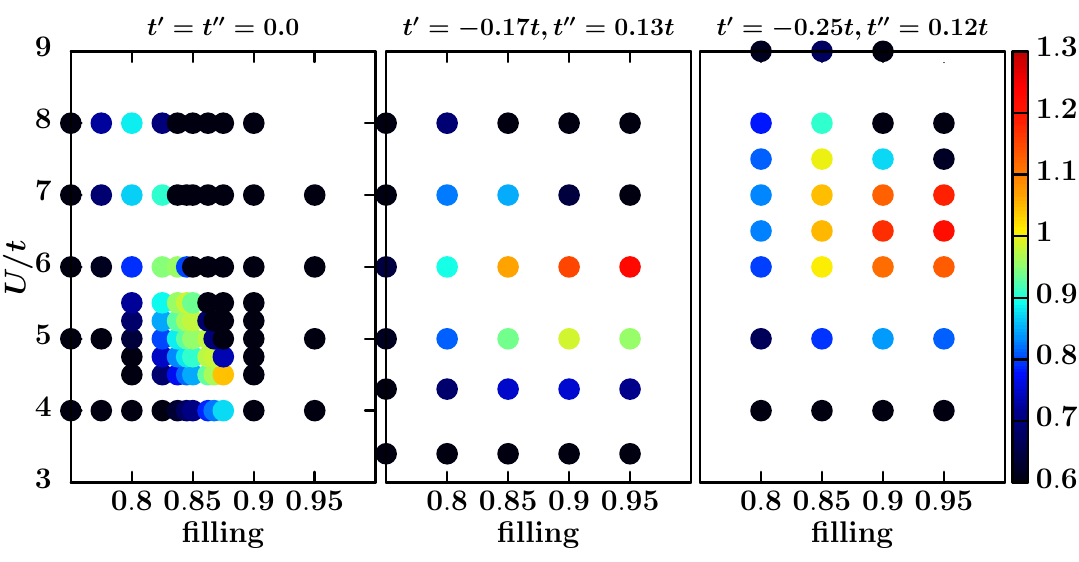}
        \caption{Directly calculated points for producing Fig. 3 in the main text. The color corresponds to the superconducting eigenvalue $\lambda$ at fixed temperature $T=0.01t$.}
        \label{Fig:calculatedpoints}
\end{figure}

In Fig.~\ref{Fig:calculatedpoints}, we display the actually calculated data points. From these points, we obtain the contour plots of Fig.~3 in the main text. For the $t^{\prime}=t^{\prime\prime}=0$ case, we linearly interpolated some irrelevant (i.e., $\lambda \le 0.6$) points: $(U/t, n)=(4.75,\; 0.775),\;(5.25,\; 0.775),\;(5.5,\; 0.75),\;(5.5,\; 0.775),\;\\(5.5,\; 0.8675)$ before making the contour plot.

\begin{figure}[tbp]
        \centering
                \includegraphics[width=\linewidth,angle=0]{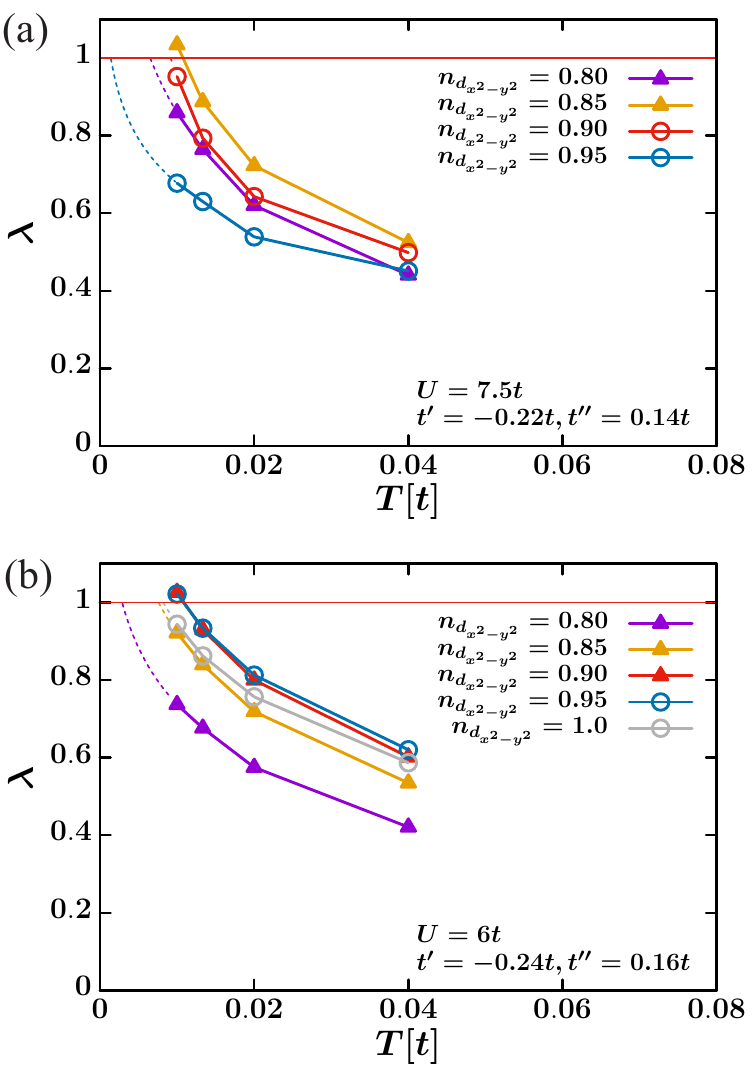}
        \caption{Temperature dependence of the superconductivity eigenvalue $\lambda$ for (a) $U=7.5t, t^{\prime}=-0.22t, t^{\prime\prime}=0.14t$ and (b) $U=6t, t^{\prime}=-0.24t,t^{\prime}=0.16t$. The extrapolation to $\lambda=1$ yields $T_c$.}
        \label{Fig:lambda-T}
\end{figure}

As for the $T_{\rm c}$ calculation, we calculate the superconducting $\lambda$ down to $T=t/100$, and extrapolated the $\lambda$ vs. $T$ curve (using the form $\lambda\approx a-b\ln(T)$) as in previous publications \cite{Kitatani2020,Kitatani2022}. As examples, we show  in Fig.~\ref{Fig:lambda-T} D$\Gamma$A results, fit function, and $T_c$ for (a) $U=7.5t,t^{\prime}=-0.22t,t^{\prime\prime}=0.14t$ and (b) $U=6t,t^{\prime}=-0.24,t^{\prime\prime}=0.16t$ results (corresponding the RbSr$_2$PdO$_3$ and $A^{\prime}_2$PdO$_2$Cl$_2$, resepctively).

Besides the superconducting eigenvalue, we also calculate the antiferromagnetic (AFM) susceptibility $\chi_{\rm sp}(Q_{\rm max},\omega=0)$. D$\Gamma$A respects constraints of the two-dimensionality (i.e., Mermin-Wagner theorem \cite{Hohenberg1967,Mermin1966}) and does not exhibit an antiferromagnetically ordered phase at finite temperatures ($\chi_{\rm sp}$ remains finite). On the other hand, we can naturally expect that AFM will stabilize for a strongly fluctuating regime (dark red region in Fig.~3 of the main text) once allowing for a weak three-dimensionality, that are present in the actual materials.

\section{VI.~ Wannier projections and consistency between DFT and Wannier bands.}

\begin{figure*}
\centering
\includegraphics[width=0.8\textwidth]{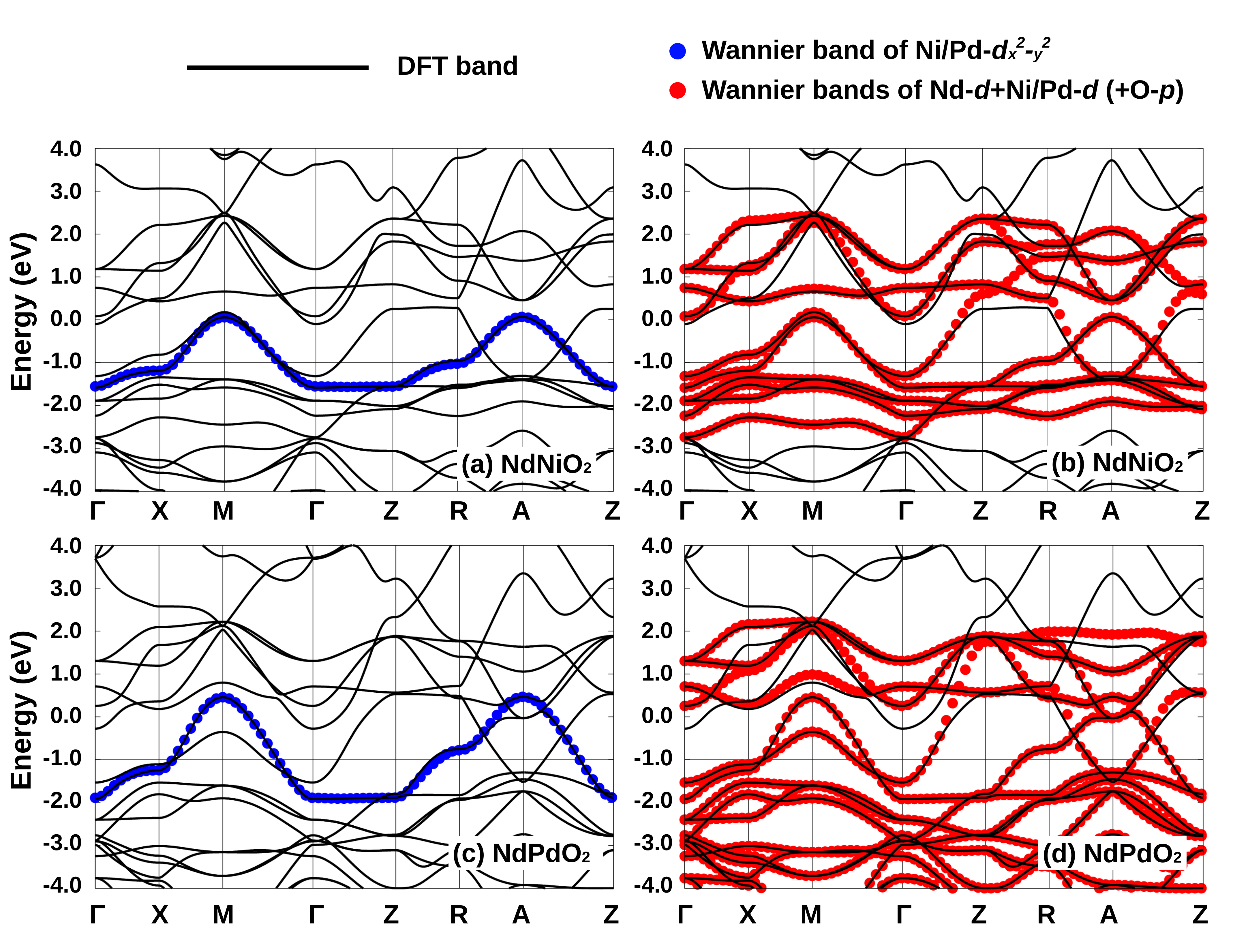}
\caption{DFT band structure of (a,b) NdNiO$_2$  and (c,d) NdPdO$_2$. The blue and red dots represent the Wannier bands of projections onto only the Pd-$d_{x^2-y^2}$ orbital and the full  Pd-4$d$ plus Nd $5d$ orbitals, respectively. In the Wannier projection for NdPdO$_2$ in (d), further the O-$p$ orbitals are  included in the projection because of the stronger Pd-$d$ - O-$p$ hybridization, compared with that of NdNiO$_2$.}
\label{Fig5_wannierNiPd}
\end{figure*}

\begin{figure*}
\centering
\includegraphics[width=0.8\textwidth]{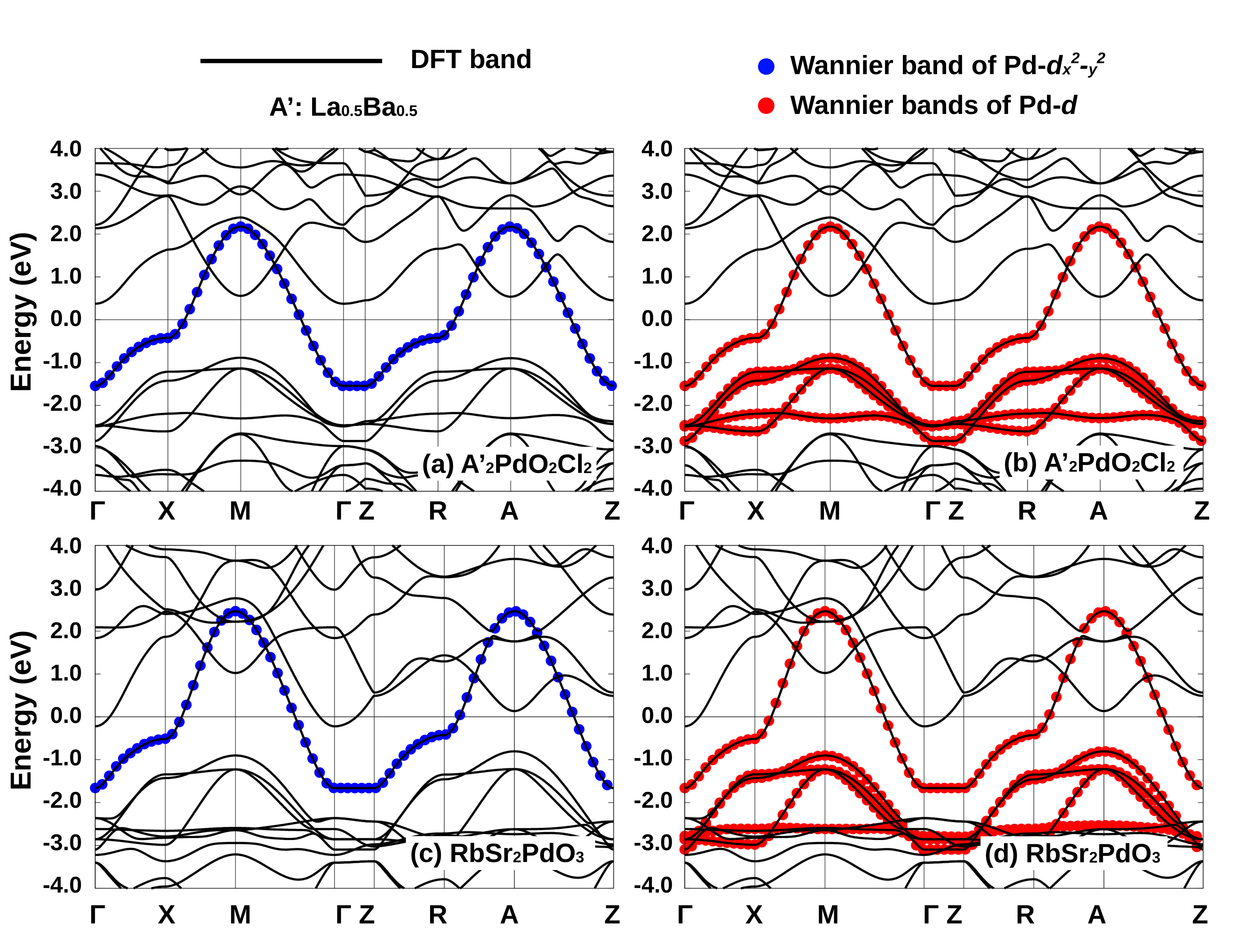}
\caption{DFT band structure of (a,b) $A'_2$PdO$_2$Cl$_2$  and (c,d) RbSr$_2$PdO$_3$. The blue and red dots represent the Wannier bands of the projection onto  (a,c) only the Pd-$d_{x^2-y^2}$ and (b,d) onto all Pd-4$d$ bands, respectively.}
\label{Fig6_wannierBaRb}
\end{figure*}

Essential for interfacing DFT and DMFT is the Wannier projections. It  not only provides the electron hopping parameters for the construction of the minimal low-energy model, but also allows  to investigate the degree of localization of the orbitals. An indication for the quality of Wannier projection is its  consistency with the target DFT bands. As in previous theoretical studies in the case of  nickelates \cite{Kitatani2020,Karp2020}, 
we performed projections for both the single $d_{x^2-y^2}$ band, as well as,  all Pd-$4d$ orbitals. In the latter case also additional orbitals that are in the vicinity of, or overlapping with, the Pd bands are included.
The single-band $d_{x^2-y^2}$ projection is here used to construct the singe-band Hamiltonian for the subsequent D$\Gamma$A calculation, while the multi-orbital ones are used for DFT+DMFT.

The corresponding hopping parameters of the single-orbital projection and
the comparison with the original DFT bands is shown in Fig.~\ref{Fig5_wannierNiPd}(a,c), Fig.~\ref{Fig6_wannierBaRb}(a,c) and Table~\ref{Table_hopping}. All these single-band projections for the Ni-$d_{x^2-y^2}$ and the Pd-$d_{x^2-y^2}$ orbital exhibit excellent agreement between the DFT and Wannier band, indicating the quality of these projections and reliability of the corresponding parameters.

For  obtaining orbital occupation and spectra, including the positions of pockets, we have done  multi-band DMFT calculations, based on the multi-band projections. For the previous calculations of NdNiO$_2$, a Nd(La)-$d$+Ni-$d$ basis with 10-bands was adopted. 
\footnote{Other multi-band projections, use Nd interstitial-$s$ instead of Nd $5d$ \cite{Nomura2019}. This interstitial-$s$ orbital is formed by a complex hybridization between Nd-$f$, Nd-$d$ and Nd(La)-$d$ orbital], or Ni-$d$+pocket-bands \cite{Lechermann2019,Lechermann2020}.}
For NdPdO$_2$, a projection onto 16 bands, 
Nd(La)-$d$+Ni-$d$+O-$p$, was necessary because of the stronger Pd-$d$-O-$p$ hybridization than in NdNiO$_2$, see Fig.~\ref{Fig5_wannierNiPd}(b,d). For $A'_2$PdO$_2$Cl$_2$ and RbSr$_2$PdO$_3$, on the other hand, a projection onto only the Pd-$d$ 5 bands is sufficient, because the other occupied bands are well separated from these  Pd-$d$ bands, see   Fig.~\ref{Fig6_wannierBaRb}(b,d). All  Wannier bands reproduce the DFT bands well, especially the low energy parts near Fermi energy.

\section{VII.~Computational details of the virtual crystal approximation for $A'_2\mathrm{PdO}_2\mathrm{Cl}_2$}

\begin{figure*}
\centering
\includegraphics[width=0.8\textwidth]{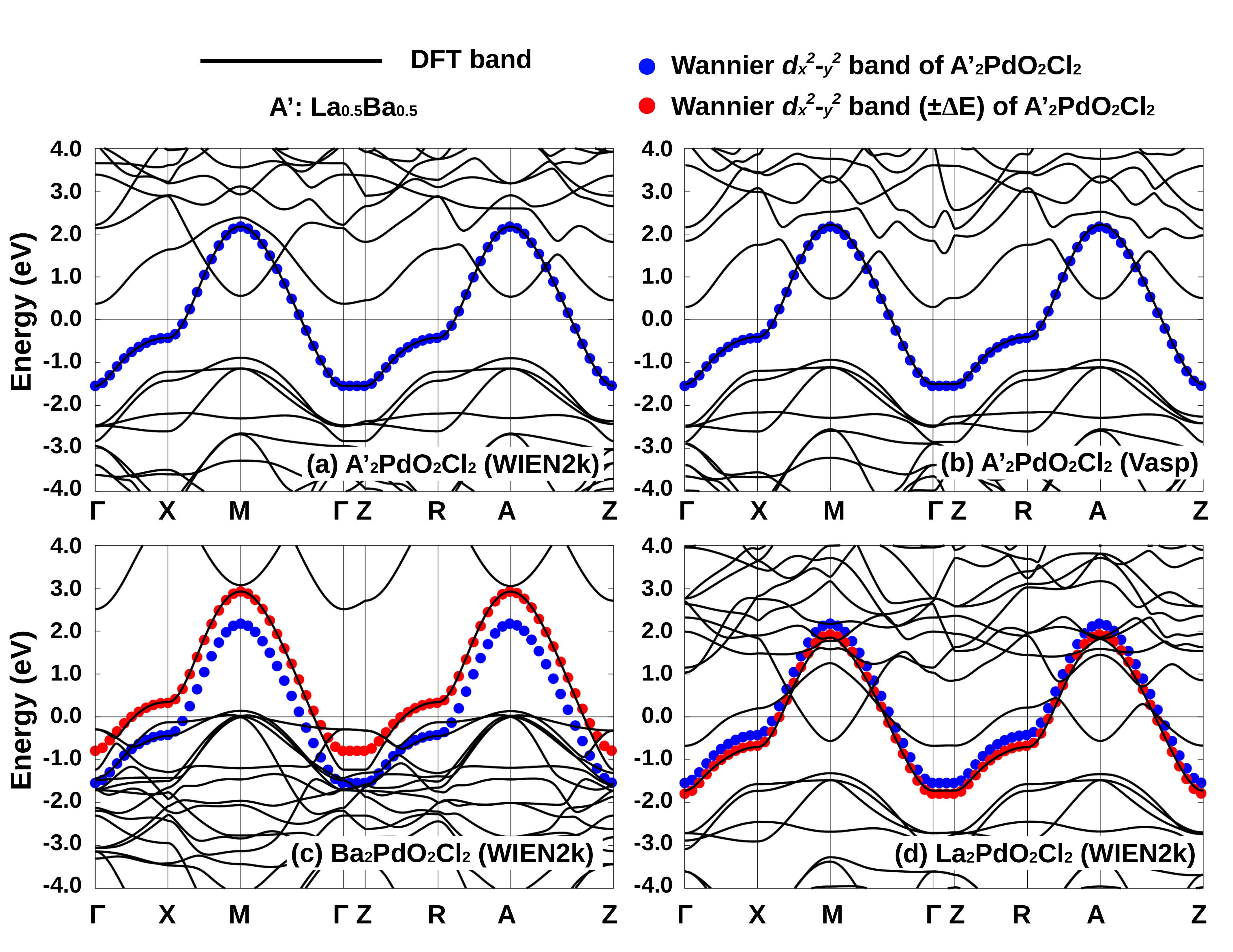}
\caption{DFT band structure of $A'^2$PdO$_2$Cl$_2$ ($A'$=La$_{0.5}$Ba$_{0.5}$) 
in the virtual crystal approximation comparing
(a) \textsc{WIEN2K} (simplified version, see text)   and (b) \textsc{Vasp}. The blue dots are the  Wannier band for the Pd-$d_{x^2-y^2}$ orbital as calculated by the \textsc{WIEN2K} Wannier projection. (c) and (d) DFT band structure of Ba$_2$PdO$_2$Cl$_2$ 
and  La$_2$PdO$_2$Cl$_2$, respectively,  as calculated in
\textsc{WIEN2K}. The Wannier band of the Pd-$d_{x^2-y^2}$ orbital in (a) is also shown (blue dots). The red dots are a constant energy shift of these by (c) $+0.75$\,eV   and (d) $-0.25$\,eV.}
\label{Fig3_vca}
\end{figure*}

The virtual crystal approximation (VCA) \cite{PhysRevB.61.7877} has been implemented in the \textsc{Vasp} code and applied to the study of Sn$_x$Ge$_{1-x}$ alloys \cite{PhysRevB.89.165201}. Such application of VCA in \textsc{Vasp} is justified because both VCA and \textsc{Vasp} are based on (Vanderbilt) pseudopotentials. However, because \textsc{WIEN2K} uses the full-potential augmented plane-wave and local-orbitals basis set to solve the Kohn-Sham equations, this full VCA is not applicable in \textsc{WIEN2K}. An alternative way to realize atomic substitution and charge doping effect in \textsc{WIEN2K} is to modify the core and valence electron states in the input configuration files 
\cite{VCAWIEN2k}.

We perform VCA calculations for $A'_2$PdO$_2$Cl$_2$ using  \textsc{WIEN2K} and \textsc{Vasp} as outlined in the last paragraph.  The corresponding bands are shown in Fig.~\ref{Fig3_vca}(a) and (b), respectively. As the comparison shows, both codes give exactly the same bands, especially the parts below $\sim$2.0\,eV.

In Fig.~\ref{Fig3_vca}(a), we additionally show the Wannier band of the Pd $d_{x^2-y^2}$ orbital. The excellent agreement between DFT and Wannier band indicates the high quality of our Wannier projection and localization of the Pd-$d_{x^2-y^2}$ orbital. In Fig.~\ref{Fig3_vca}(c,d), we compute and show the DFT band of (c) undoped Ba$_2$PdO$_2$Cl$_2$ and (d) La$_2$PdO$_2$Cl$_2$. In the former and later cases the bivalent Ba$^{2+}$ and trivalent La$^{3+}$ lead to nominal Pd$^{2+}$ (4$d^8$) and Pd$^{0}$ (4$d^{10}$), respectively. Please note that in La$_2$PdO$_2$Cl$_2$ due to the hybridization between La-$d$ and Pd-$d$, the oxidation state of Pd is in fact between Pd$^{1+}$ and  Pd$^{0}$, as in $A'_2$PdO$_2$Cl$_2$. 

In Fig.~\ref{Fig3_vca}(c,d), we show again the Pd-$d_{x^2-y^2}$ Wannier band taken from Fig.~\ref{Fig3_vca}(a), and add (subtract) a constant, leading to the red dots in Fig.~\ref{Fig3_vca}(c,d). The shifted Wannier bands exhibit excellent consistency between the Pd-$d_{x^2-y^2}$ bands in Fig.~\ref{Fig3_vca}(c,d), indicating the atomic substitution and charge doping effects can, in fact, be well described by a rigid band shift. This explains why the simplified VCA method in \textsc{WIEN2K} yields essentially the  same results as \textsc{Vasp}.

\section{VIII.~phonon calculations and structural  distortions of $\mathbf \mathrm{LaPdO}_2$}

\begin{figure*}
\centering
\includegraphics[width=0.8\textwidth]{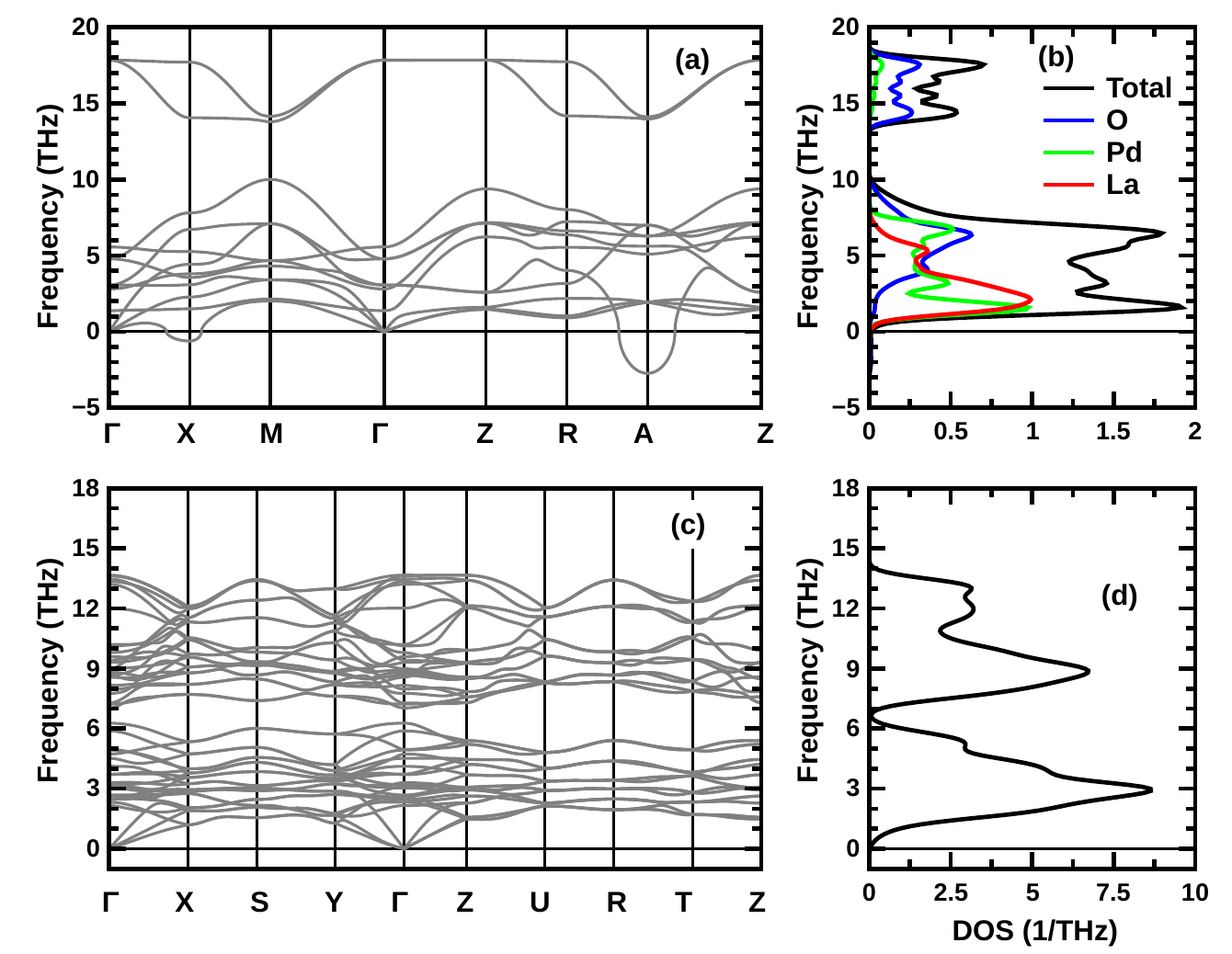}
\caption{(a,c) DFT phonon spectra and (b,d)  phonon density of states of the (fully relaxed) ideal $P4/mmm$ (a,b) and distorted $Pbcn$ (c,d) phase of LaPdO$_2$.}
\label{Fig1_phonon}
\end{figure*}

\begin{figure*}
\centering
\includegraphics[width=0.8\textwidth]{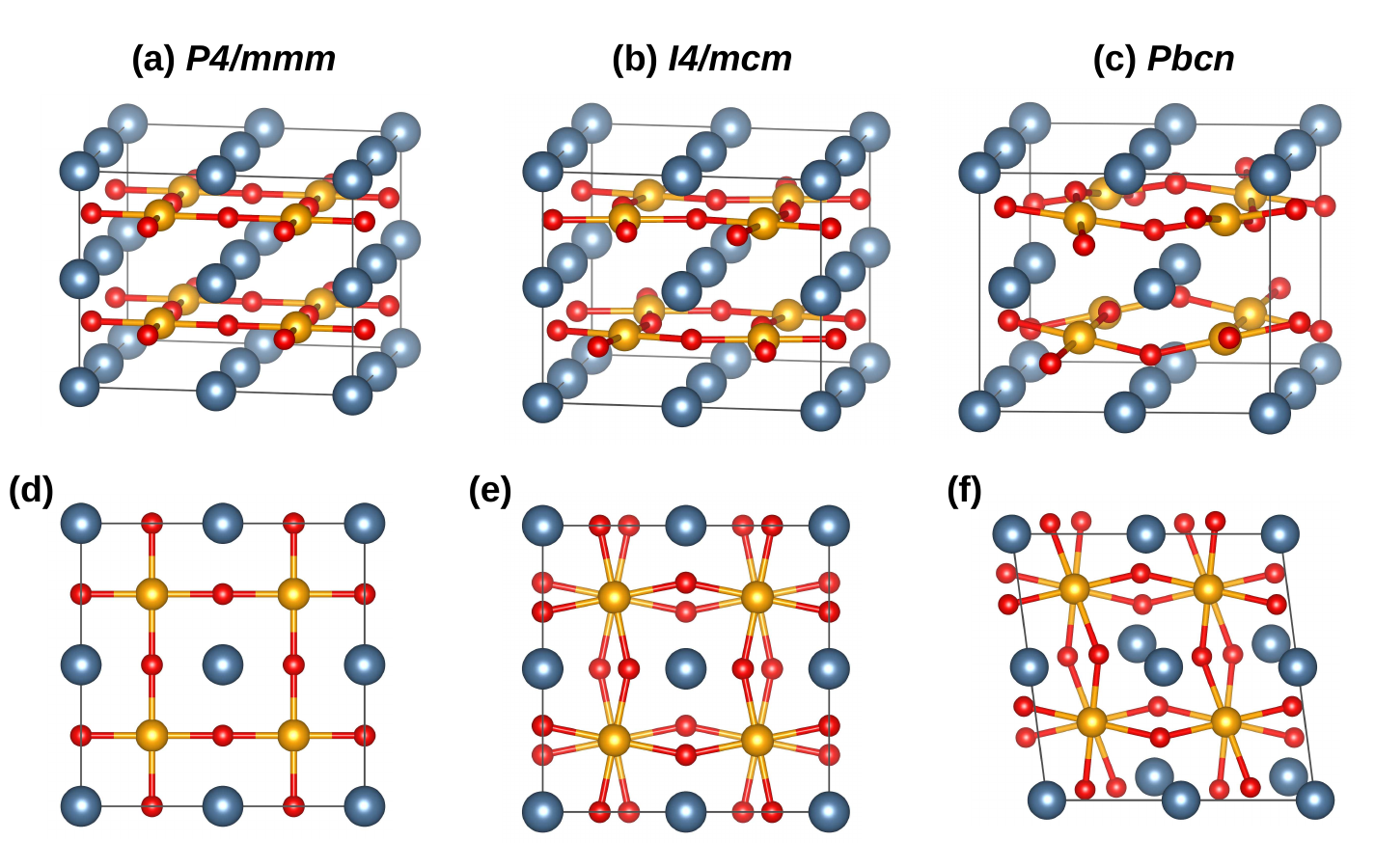}
\caption{DFT-relaxed structures of LaPdO$_2$: (a,d)  ideal tetragonal $P4/mmm$ phase; (b,e) tetragonal $I4/mcm$ phase; (c,f) orthorhombic $Pbcn$ phase. The panels (d,e,f) are the top-view along the  $z$-direction (001) of panels (a,b,c), respectively.}
\label{Fig2_structure}
\end{figure*}

To investigate the dynamical stability of NdPdO$_2$, we compute its phonon spectra  with the frozen phonon method (finite displacement method) using the \textsc{Phonony} \cite{togo2015first} code interfaced with \textsc{Vasp} \cite{kresse1996efficiency,PhysRevB.54.11169}. For the phonon spectra of the other two palladates, $A'_2$PdO$_2$Cl$_2$ and RbSr$_2$Pd$_2$O$_3$, similar calcualtions can be found in  Fig.~11(b,c) of Ref.~\cite{Motoaki2019}. Here, to increase computational efficiency and to avoid the issue of bad convergence arising from Nd-4$f$ bands, we replace Nd by La. Previous theoretical research \cite{Botana2019} and the experimentally confirmation of superconductivity in both Ca-doped LaNiO$_2$ \cite{Zeng2021} and Sr-doped NdNiO$_2$ \cite{zeng2020,Li2020} indicate that the replacement of Nd by La does not significantly change the physics. $T_{\rm c}$ is reduced a bit in the La compound, which likely originates mainly from different levels of disorder. The corresponding phonon spectra of the ideal $P4/mmm$ phase of LaPdO$_2$ is shown in Fig.~\ref{Fig1_phonon} and the structural evolution is shown in Fig.~\ref{Fig2_structure}(a-c).

We first check the phonon spectra of the ideal $P4/mmm$ phase LaPdO$_2$. Using the PBESol functional, the optimized lattice parameters are $a$=$b$=4.126\,\AA, $c$=3.265\,\AA. Compared with La(Nd)NiO$_2$, the $a$ ($c$) lattice parameter is increased (reduced) by 6.1\% (2.1\%). The instability of $P4/mmm$ LaPdO$_2$ is evidenced by negative frequencies in its phonon spectra [Fig.~\ref{Fig1_phonon}(a,b)]. The phonon spectra and atomic-resolved DOS are computed by employing a 2$\times$2$\times$2 supercell of LaPdO$_2$, which is shown in Fig.~\ref{Fig2_structure}(a). As shown in Fig.~\ref{Fig1_phonon}, the phonon spectrum of $P4/mmm$ LaPdO$_2$ reveals a dynamical instability at both the $X$ [$q$=(1/2,0,0)] and $A$ [$q$=1/2,1/2,1/2] points of the $q$-path. These unstable phonons are in fact consistent with those in SmNiO$_2$, YNiO$_2$ \cite{PhysRevMaterials.6.044807}, LuNiO$_2$ \cite{subedi2022possible,PhysRevB.105.115134}, and EuNiO$_2$  \cite{PhysRevB.105.115134}. We further predict the structures with lower symmetry and total energies on the basis of the eigenvectors (atomic vibrations) induced by the $X_{2}^{-}$, $A_{3}^{+}$ mode and the combination between both of them, using group theory and the workflow of \cite{subedi2022possible}.
The eigenvector of the unstable phonon at $X$-point is $X_{2}^{-}$ \cite{subedi2022possible} and is related to a ferroelectric distortion perpendicular to the PdO$_2$ layers. The phonon mode at the $A$-point is $A_{3}^{+}$ and is related to the in-plane rotation of PdO$_4$ planers, leading to a $I4/mcm$ symmetry \cite{PhysRevB.105.115134} [Fig.~\ref{Fig2_structure}(b)] phase. Further, the combination of $X_{2}^{-}$ and $A_{3}^{+}$ corresponds to an orthorhombic distortion, i.e., the $Pbcn$ phase [Fig.~\ref{Fig2_structure}(c)]. After (non-spin-polarized) DFT-PBESol structural relaxations, the $I4/mcm$ and $Pbcn$ phase are energetically more stable than the ideal $P4/mmm$ phase by 102\,meV and 225\,meV per LaPdO$_2$ formula unit, respectively. We hence conclude that the $P4/mmm$ ($Pbcn$) structures are corresponding to high (low) temperature phases. 

We additionally compute the Pd-$d_{x^2-y^2}$ hopping of the ideal $P4/mmm$, tetragonal $I4/mcm$ and the low symmetry $Pbcn$ phase of LaPdO$_2$.  As shown in Table~\ref{Table_hopping}, the 1st nearest hopping $t$ is reduced from -537\,meV in ideal $P4/mmm$ phase to -357\,meV in $I4/mcm$ phase, finally to -304\,meV in $Pbcn$ phase.
Due to the both, the reduced $t$ and $t^\prime$, the  DFT bandwidth is reduced is reduced from $\sim$4.4\,eV in the $P4/mmm$ phase to $\sim$2.4\,eV in the $Pbcn$ phase.
The $t$ in $I4/mcm$ and $Pbcn$ phase is even smaller than that in NdNiO$_2$ (-397\,meV), indicating stronger correlations than in the ideal $P4/mmm$ phase.

\begin{figure*}
\centering
\includegraphics[width=0.95\textwidth]{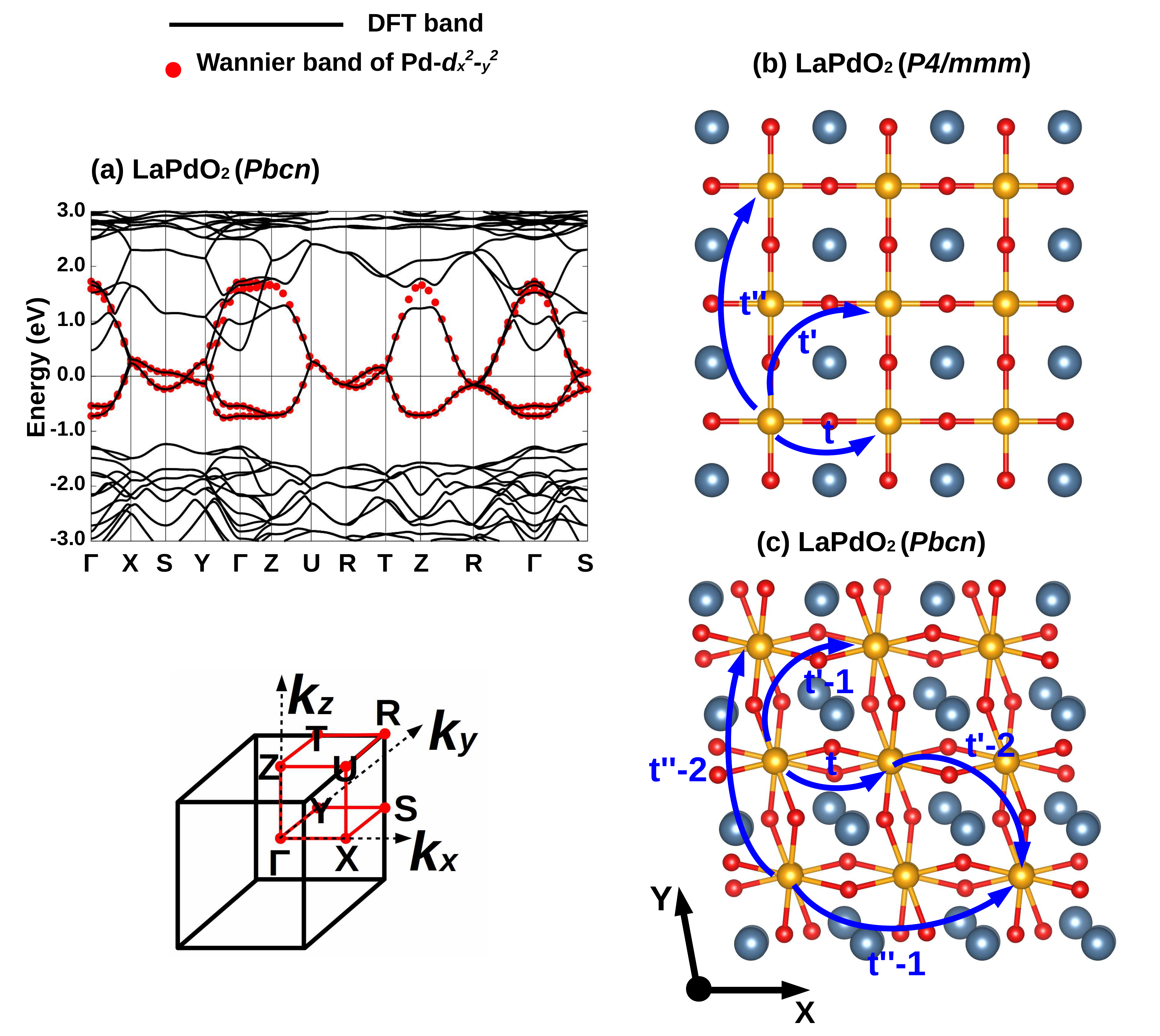}
\caption{(a) Comparison between DFT and Wannier tight-binding Hamiltonian bands of distorted $Pbcn$ phase of LaPdO$_2$, the bottom-left panel shows the brillouin zone. Major hopping terms of (b) ideal $P4/mmm$ and (c) distorted $Pbcn$ phase of LaPdO$_2$. The $t$, $t^{\prime}$ and  $t^{\prime\prime}$ indicates the 1st, 2nd and 3rd nearest hopping between Pd-$d_{x^2-y^2}$ orbital, obtained from Wannier projection calculations. Please note that in the distorted $Pbcn$ phase of LaPdO$_2$, both the $t^{\prime}$ and  $t^{\prime\prime}$ are anisotropic due to the orthorhombic distortion, i.e., the hoppings along (110) and ($\bar{1}$10), as well as along the (200) and (020) direction, are different. The $t^{\prime}$-1 (-2) and  $t^{\prime\prime}$-1 (-2) corresponds to the 1st (2nd) row value of $Pbcn$ phase in Table.~\ref{Table_hopping}.}
\label{Fig5_single}
\end{figure*}

For the comparison between DFT and Wannier bands of distorted $Pbcn$ phase of LaPdO$_2$, see Fig.~\ref{Fig5_single}. Compared with the bands of idea $P4/mmm$ phase, the bandwidth of $Pbcn$ phase of LaPdO$_2$ is effectively reduced from 4.60\,eV in $P4/mmm$ phase to 2.45\,eV in $Pbcn$ phase, these are consistent with the reduced major hoppings shown in Table~\ref{Table_hopping}. Additionally, in both Table~\ref{Table_hopping} and Fig.~\ref{Fig5_single}(b,c) we demonstrate the emergence of anisotropic hoppings terms of $t^\prime$ and $t^{\prime\prime}$ along (110) and ($\bar{1}$10), as well as along the (200) and (020) direction, respectively. Such anisotropic hopping is a consequence of the orthorhombic distortion is $Pbcn$ phase.

\begin{table}[tb]
\caption{Wannier hoppings of the Pd $d_{x^2-y^2}$ orbital for the ideal $P4/mmm$, tetragonal $I4/mcm$ and orthorhombic $Pbcn$ phase of LaPdO$_2$ (in unit of meV). The $t$, $t^{\prime}$ and $t^{\prime\prime}$ indicate the 1st, 2nd and 3rd nearest hopping along the real-space vectors of (100), (110) and (200), respectively. The ratios between $t^{\prime}/t$ and $t^{\prime\prime}/t$ are also given. The emergence of the second $Pbcn$ values is due to the orthorhombic distortion in $Pbcn$ phase, leading to nonequivalent $t^{\prime}$ and $t^{\prime\prime}$ along (110) and ($\bar{1}$10) as well as between the (200) and (020) direction.}
\label{Table_hopping}
\centering
\begin{tabular}{c|c|c|c|c|c}
\hline 
\hline
Phase & $t$ & $t'$ & $t''$ & $t^{\prime}/t$ & $t^{\prime\prime}/t$ \\
\hline
$P4/mmm$ & -537 & 97 & -71 & -0.181 & 0.131 \\
\hline
$I4/mcm$ & -357 & 77 & -15 & -0.214 & 0.043 \\
\hline
$Pbcn$   & -304 & 56 &  3 & -0.184 & 0.014 \\
    & -304 & 95 &  -4 & -0.313 & -0.012 \\
\hline
\hline
\end{tabular}
\end{table}

\section{IX.~DFT+$U$ magnetic total energies of $\mathrm A'_2\mathrm{PdO}_2\mathrm{Cl}_2$}

\begin{figure*}
\centering
\includegraphics[width=0.95\textwidth]{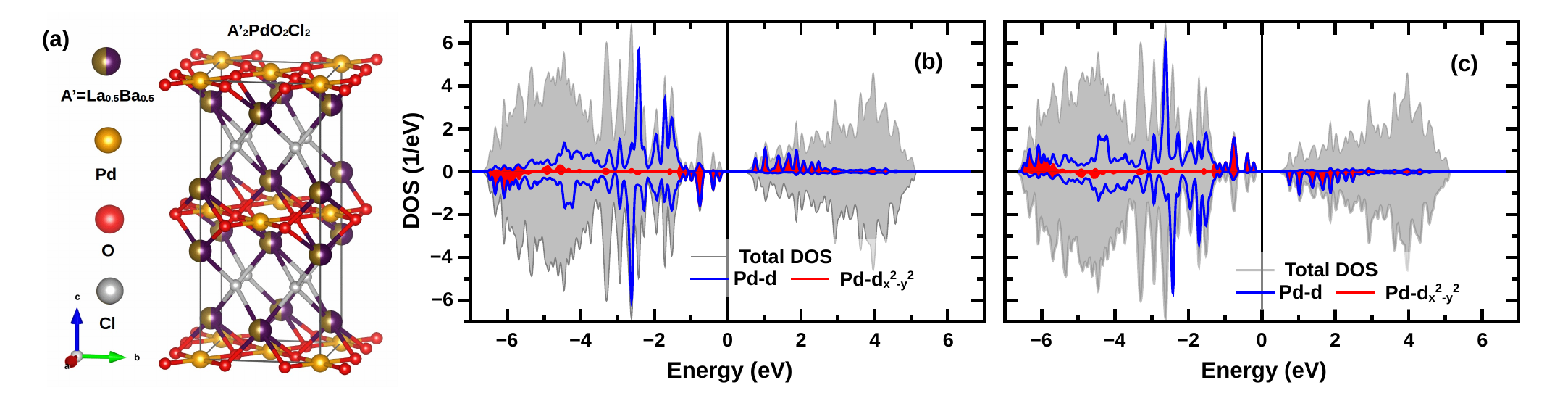}
\caption{(a) DFT crystal structure of  $A'_2$PdO$_2$Cl$_2$  and (b,c)  density-of-states of antiferromagentic DFT+$U$ at $U$=3.3\,eV for the two inequivalent Pd sites of the supercell of (a) and $G$-type antiferromagentism. The upper and lower panels of (b,c) denote the spin-up and -down channels respectively.  }
\label{Fig4_afm}
\end{figure*}

To investigate the magnetic ground state of $A'_2$PdO$_2$Cl$_2$ ($A$=La$_{0.5}$Ba$_{0.5}$), we perform DFT+$U$ ($U$=3\,eV, 3.3\,eV and 4\,eV) calculations with a $\sqrt{2}\times$$\sqrt{2}$$\times$1 supercell, as shown in Fig.~\ref{Fig4_afm}(a). In each layer of (PdO$_2$)$_2$, there are two Pd sites, and in total two such (PdO$_2$)$_2$ layers in the supercell. We define the following three types of magnetic order: FM (inter-layer FM and intra-layer FM), $A$-AFM (inter-layer AFM and intra-layer FM), $G$-AFM (inter-layer AFM and intra-layer AFM) \footnote{Due to the shift between the 1st and 2nd (PdO$_2$)$_2$ layers, the Pd of the next layer is in-between the two Pd sites of the previous layer in  $A'_2$PdO$_2$Cl$_2$. Hence, $C$-type AFM order (inter-layer FM and intra-layer AFM) is not possible in $A'_2$PdO$_2$Cl$_2$.}. Our DFT+$U$ calculations confirm $G$-AFM to be the magnetic ground state for all $U$ values. Specifically, the $G$-AFM order is for the aforementioned three $U$ values, more stable than FM order by 84\,meV, 103\,meV and 162\,meV per $A'_2$PdO$_2$Cl$_2$ chemical formula. The magnetic moments in FM state for all three considered $U$ values are $\sim 0.2$\,$\mu_B$, $0.25$\,\,$\mu_B$ and $0.40$\,$\mu_B$ per Pd, such small moments are not very effective in reducing their total energy. As a consequence, the total energy of FM states are only $<$1\,meV ($U=3$\,eV), $\sim$2\,meV ($U=3.3$\,eV) and $\sim$2\,meV ($U=4$\,eV) lower than those of PM states (per $A'_2$PdO$_2$Cl$_2$ chemical formula). 
We find the energies obtained from $A$-AFM setup calculations are the same as for PM order because they converge to PM state, indicating the inter-layer (anti-ferromagnetic) couplings/interactions between PdO$_2$ layers are negligible and too weak to stabilize $A$-AFM order. This can be explained by the long distance between PdO$_2$ layers along the $z$-direction ($\sim$7.32\,\AA\ according to DFT-PBESol structural relaxations). 

Fig.~\ref{Fig4_afm}(b,c) show the density-of-states (DOS) of $G$-type antiferromagnetic order  in $A'_2$PdO$_2$Cl$_2$, resolved for  the two inequivalent sites and spin-up and spin-down channel ($U$=3.3\,eV as in our  multi-orbital DMFT calculations). With such a degree of correlations a gap of $\sim$0.6\,eV is opened in the presence of $G$-AFM order. The highest occupied orbital is Pd-$d_{x^2-y^2}$, indicting a single-band AFM orders that is similar to cuprates. The magnetic moment projected onto the Pd site is 0.55\,$\mu_B$ ($U$=3\,eV), 0.57\,$\mu_B$ ($U$=3.3\,eV) and 0.63\,$\mu_B$ ($U$=4\,eV) per Pd, residual magnetic moments are located at the O-$p$ orbital and interstitially. 
This demonstrates the $d$-$p$ hybridization between Pd and O is stronger than that in nickelates. This puts palladates in-between cuprates and nickelates, as already indicated from the energy levels in Fig.~1 of the main text. 

\bibliography{full}